\documentclass[12pt,draftclsnofoot,journal,onecolumn]{IEEEtran}
\usepackage{graphicx}
\usepackage{subfigure}
\usepackage{color}
\usepackage{multicol}
\usepackage{amssymb}
\usepackage{multirow}
\usepackage{amsmath}
\usepackage{bm}
\usepackage{cite}
\usepackage{gensymb}
\usepackage{amsfonts}
\usepackage{mathrsfs}
\usepackage{amsmath}
\usepackage{algorithm}
\usepackage{algorithmic}
\usepackage{amsthm}

\usepackage{tabularx}
\usepackage{balance}
\usepackage{mathrsfs}
\usepackage{array}
\usepackage{multirow}
\newcommand{\PreserveBackslash}[1]{\let\temp=\\#1\let\\=\temp}
\newcolumntype{C}[1]{>{\PreserveBackslash\centering}p{#1}}
\newcolumntype{L}[1]{>{\PreserveBackslash\raggedright}p{#1}}
\newcolumntype{R}[1]{>{\PreserveBackslash\raggedleft}p{#1}}
\usepackage[flushleft]{threeparttable}
\newcommand{\mb}[1]{{  \mathbf  #1}}  %\mathbf  \bm

\newtheorem{thm}{Lemma}
\begin{document}
	
	\bibliographystyle{IEEEtran} % use IEEEtran.bst style
	
	\title{Near-Field Rainbow: Wideband Beam Training for XL-MIMO}
	\author{\IEEEauthorblockN{ 
%			Mingyao Cui, \emph{Student Member, IEEE}, Linglong Dai, \emph{Fellow, IEEE}, Zhaocheng Wang, \emph{Fellow, IEEE}, and Ning Ge, \emph{Member, IEEE}
Mingyao Cui,  Linglong Dai,  Zhaocheng Wang,  Shidong Zhou, and Ning Ge
		}
	\thanks{
				All authors are with the Beijing National Research Center for Information Science and Technology (BNRist) as well as the Department of Electronic Engineering, Tsinghua University, Beijing 100084, China (e-mails: cmy20@mails.tsinghua.edu.cn;  daill@tsinghua.edu.cn; 
		zcwang@tsinghua.edu.cn; zhousd@tsinghua.edu.cn; 
		gening@tsinghua.edu.cn).
		
		This work was supported in part by the National Key Research and Development Program of China (Grant No.2020YFB1807201), in part by the National Natural Science Foundation of China (Grant No. 62031019), and in part by the European Commission through the H2020-MSCA-ITN META WIRELESS Research Project under Grant 956256. 
%		(\emph{Corresponding author: Linglong Dai.})
}
}
	\maketitle
	\IEEEpeerreviewmaketitle
	\begin{abstract}
Wideband extremely large-scale multiple-input-multiple-output (XL-MIMO) is a 
promising technique to achieve Tbps data rates in future 6G systems through 
beamforming and spatial multiplexing.
Due to the extensive bandwidth and the huge number of antennas for wideband 
XL-MIMO, a significant near-field beam split effect will be induced, where 
beams at different frequencies are focused on different locations. The 
near-field beam split effect results in a severe array gain loss, so existing 
works mainly focus on compensating for this loss by utilizing the time delay 
(TD) beamformer. By contrast, this paper demonstrates that although the 
near-field beam split effect degrades the array gain, it also provides a new 
possibility to realize fast near-field beam training. 
{\color{black}Specifically, we first 
reveal the mechanism of the near-field controllable beam split effect. This 
effect indicates that, by dedicatedly designing the delay parameters, a TD 
beamformer is able to control the degree of the near-field beam split effect, 
i.e., beams at different frequencies can flexibly occupy the desired location 
range. Due to the similarity with the dispersion of natural light caused by a 
prism, this effect is also termed as the near-field rainbow in this paper. 
Then, 
taking advantage of the near-field rainbow effect, a fast wideband beam 
training scheme is proposed. 
In our scheme, the close form of the beamforming 
vector is elaborately derived to enable beams at different frequencies to be 
focused on different desired locations. By this means, the optimal beamforming 
vector with the largest array gain can be rapidly searched out by generating 
multiple beams focused on multiple locations simultaneously through only one 
radio-frequency (RF) chain. 
}Finally, simulation results demonstrate the proposed scheme is able to realize 
near-optimal near-field beam training with a very low training overhead.
	\end{abstract}
	
	\begin{IEEEkeywords}
		XL-MIMO, near-field, wideband, beam training.\vspace{-3mm}
	\end{IEEEkeywords}	
	\section{Introduction}
	Wideband extremely large-scale multiple-input-multiple-output (XL-MIMO) has been regarded as a promising technology to meet the capacity requirement for future 6G communications \cite{XLMIMO_Carvalho2020, radiostrip_Frenger2017}. 
	Benefiting from the huge spatial multiplexing gain provided by a very large 
	number of antennas, XL-MIMO is able to significantly increase the spectrum 
	efficiency \cite{6Gchallenge_Rappaport2019}. Moreover, XL-MIMO can also 
	provide very high beamforming gain to compensate for the severe path loss 
	at millimeter-wave (mmWave) or terahertz (THz) band, which may provide tens 
	of GHz-wide bandwidth to enable Tbps data rates for future 6G networks 
	\cite{THzsurvey_Elayan2020}. 
	
	Nevertheless, due to the very large array aperture and the extensive 
	bandwidth at high frequencies, a significant near-field beam split effect 
	will be introduced \cite{NFBF_Cui2021}.
	Firstly, compared to the far-field propagation in conventional 5G systems 
	where electromagnetic (EM) wavefronts can be approximated as \emph{planar}, 
	the deployment of XL-MIMO, especially at high-frequency, 
	indicates that the near-field propagation
	will become essential in 6G 
	networks, where the EM wavefronts have to be accurately modeled as 
	\emph{spherical} \cite{NearCE_Cui2022,NearBF_Zhang2022}. 
	{\color{black}The boundary between the far-field and near-field regions is 
	determined by 
	the Rayleigh distance, which is proportional to the square of
	the array aperture and the signal frequency \cite{fresnel_Selvan2017}. 
	With the increased array aperture and frequency of XL-MIMO, 
	its near-field range can be several dozens and even hundreds of meters 
	\cite{NearMag_Cui2022}.
	For example, for a 1-meter diameter array 
	operating at 30 GHz, the near-field 
	region extends to distances of 200 meters, dominating the typical outdoor 
	communication environments.}
	As a result, spherical wavefronts should be exploited to realize near-field
	\textbf{beamfocusing} (near-field beamforming) in XL-MIMO systems to allow 
	focusing signals at a 
	\emph{specific location}, rather than the conventional far-field 
	\textbf{beamsteering} (far-field beamforming) that steers signals towards a 
	\emph{specific angle} \cite{NearMag_Cui2022}.

	Secondly, the very large bandwidth results in the near-field beam split 
	effect. 
In XL-MIMO systems, phase-shifter (PS) based beamformer is widely considered 
to generate focused beams aligned with certain locations to provide 
beamfocusing gain \cite{OMP_Ayach2014}. 
	Such a beamformer works well for narrowband systems. 
	{\color{black}However, for wideband systems, the beams at different 
	frequencies with spherical 
	wavefronts will be 
	focused on different physical locations due to the use of nearly
	\emph{frequency-independent} PSs, which is referred to as 
	the \textbf{near-field beam split} effect \cite{NFBF_Cui2021}.  }
	This effect results in a severe array gain loss since the beams over 
	different frequencies cannot be aligned with the target user in a certain 
	location, which should be carefully addressed.
%	\emph{frequency-independent} beamforming. 
%	Since the practical wideband channel is \emph{frequency-dependent}, the 
%generated beams with near-field spherical wavefronts over different 
%frequencies 
%may split into different spatial angles and distances, 

	\subsection{Prior Works}
As the antenna number is not very large for current 5G massive MIMO 
communications \cite{5Gparadigm_Mumtaz2016}, existing works mainly focus on a 
simplified \emph{far-field beam split} effect. To be specific, the channels are 
modeled under the planar wavefronts, and the beams over different frequencies 
are split into different spatial angles, where the distance information is 
ignored.
	These works can be generally classified into two categories, i.e., 
	techniques overcoming the far-field beam split effect 
	\cite{TTDhardware_Hashemi2008, TTD_Gha2019, TTD_Lin2017, DPP_Tan2019, 
	DPP_Dang2021} and those taking advantage of it \cite{ TTDBT_Yan2019, 
	TTDBT_Bol2021, THzBT_Tan2021}.
	The first category desires to mitigate the array gain loss caused by the 
	far-field beam split. 
	To realize this target, introducing time-delay circuits into beamforming 
	structures, such as true-time-delay array \cite{TTDhardware_Hashemi2008, 
	TTD_Gha2019} or delay-phase precoding structure \cite{DPP_Tan2019, 
	DPP_Dang2021}, is considered to be promising. Thanks to the 
	\emph{frequency-dependent} phase shift provided by the time-delay circuit, 
	the generated beams over the entire bandwidth can be aligned with a certain 
	spatial angle, and thus the far-field beam split effect can be alleviated.
	For the second category of taking advantage of the far-field beam split, it 
	has been proved that time-delay circuits can not only mitigate the 
	far-field 
	beam split, but also flexibly control its degree \cite{TTDBT_Yan2019, 
	THzBT_Tan2021, TTDBT_Bol2021}. By carefully designing the delay parameters, 
	the covered angular range of the beams over different frequencies can be 
	controlled. 
	Benefiting from this fact, very fast channel station information (CSI) 
	acquisition in the far-field, such as fast beam training or beam tracking, 
	can be realized. 
%	Specifically, beam training and beam tracking that directly estimates the physical angles of far-field channel paths are widely considered to obtain the CSI. 
	In  \cite{TTDBT_Yan2019, TTDBT_Bol2021}, based on the true-time-delay array architecture, fast beam training schemes with only one pilot overhead were studied by generating frequency-dependent beams to simultaneously search multiple angles. 
	In \cite{THzBT_Tan2021}, a fast beam tracking scheme based on the delay-phase precoding architecture was proposed by adaptively adjusting the degree of the far-field beam split according to the user mobility. 
	From the discussion above, we can find that although far-field beam split 
	results in a severe beamsteering gain loss that should be addressed, it can 
	also benefit the fast far-field CSI acquisition in massive MIMO systems. 
	
	When it comes to the XL-MIMO systems, the more realistic near-field beam 
	split effect should be considered, since the antenna number is huge. 
	Recently, researchers in \cite{NFBF_Cui2021} have tried to compensate for 
	the corresponding array gain loss in the near-field range. 
	Specifically, the near-field beam split effect was defined and analyzed in 
	our previous work \cite{NFBF_Cui2021}, and the time-delay (TD) based 
	beamformer 
	was also utilized to overcome this effect. 
	We have proposed to partition the entire array into multiple sub-arrays, and then the user can be assumed to be located within the near-field range of the entire array but in the far-field range of each sub-array. Based on this partition, time-delay circuits can also be utilized to compensate for the group delays across different sub-arrays induced by near-field spherical wavefronts.
	As a result, the beams over the entire bandwidth can be focused on the desired spatial angle and distance, and the near-field beam split effect is alleviated accordingly. 
	
	Efficient design of wideband XL-MIMO beamfocusing to alleviate the 
	near-field beam split effect requires accurate near-field CSI. 
	To meet this 
	requirement, an intuitive way is to directly utilize the existing wideband 
	far-field CSI acquisition schemes \cite{TTDBT_Yan2019, THzBT_Tan2021, 
	TTDBT_Bol2021} to estimate the near-field wideband channel.  
	However,  since the 
	far-field planar wavefront mismatches the near-field spherical wavefront, 
	these methods \cite{TTDBT_Yan2019, THzBT_Tan2021, TTDBT_Bol2021} may be not 
	valid in the near-field range. 
	{\color{black}To cope with this problem, 
	inspired by the classical far-field hierarchical beam training scheme 
	\cite{BT_Noh2017}, 
	a near-field hierarchical beam training method is proposed in 
	\cite{NFBT_Hong2021}, which uniformly searches multiple angles and 
	distances to  obtain the near-field CSI. }
	Moreover, from the perspective of near-field array gain,  
	\cite{NearCE_Cui2022} proved that the distances should be non-uniformly 
	searched for improving channel estimation accuracy.
	Nevertheless, existing near-field CSI acquisition methods assume that, the bandwidth is not very large, so the near-field beam split effect is not considered. Moreover, unlike 
	the angle-dependent far-field CSI, to obtain the near-field CSI, the angle and distance information should be estimated simultaneously,
	which may result in unacceptable pilot overhead for near-field beam training \cite{NFBT_Hong2021}. 
	Unfortunately, to the best of our knowledge, determining how to obtain 
	accurate wideband XL-MIMO near-field CSI with acceptable pilot overhead has 
	not been studied in the literature. 
	
%	To realize efficient near-field 
%	
%	It is known from \cite{NFBF_Cui2021} that time-delayers can also overcome the near-field beam split effect. Then, analog to the two categories of research on far-field beam split we have discussed before, some critical questions arise naturally, ``do time-delayers can control the degree of near-field beam split, just like in the far-field scenarios?" and ``whether the near-field beam split effect can provide a new dimension to realize fast near-field CSI acquisition?". 
%	Since the near-field channel with spherical wavefront is more complex than the far-field channel with planar wavefront, these questions seem non-obvious. 
%	Moreover, unlike the angle-related-only far-field CSI, to obtain the near-field CSI, the angle and distance information should be estimated simultaneously, which may result in unacceptable pilot overhead for XL-MIMO \cite{NFBF_Cui2021}. It is indicated in \cite{NFBT_Hong2021} that the overhead for near-field beam training can be dozens of times that of far-field. Therefore, the solutions to these questions are also essential for practical XL-MIMO communications.
%	However, to the best of our knowledge, there are no related works answering these interesting and important questions.

	\subsection{Our Contributions}
	To fill in this gap, inspired by the two categories of research on the far-field beam split, i.e., overcoming and taking advantage of it, we unveil  that 
	 although the near-field beam split effect degrades the array gain, it can 
	 also provide a new possibility to benefit the near-field CSI acquisition. 
	Based on this new finding, we propose a fast wideband near-field beam training scheme by taking advantage of the near-field beam split effect. 
	Specifically, our contributions are summarized as follows.
%	Unlike existing works focus on alleviating the near-field beam split by utilizing time-delayers, we demonstrate that time-delayers can not only make the near-field beam split effect avoidable but also make it available.
\begin{itemize}
	\item  {\color{black}Firstly, we prove the effect of near-field 
	controllable 
	beam 
	split, i.e., time-delay circuits can flexibly control the degree of 
	near-field beam split.   
	Specifically, it has been proved in far-field scenarios that, time-delay 
	circuits can control the 
	covered angular range of the beams over different frequencies when the 
	distance is very large 	\cite{TTDBT_Yan2019, 
		THzBT_Tan2021, TTDBT_Bol2021}.
	By contrast, we will reveal that not only the covered angular range, but 
	also the covered distance range of the beams over different frequencies can 
	be controlled by the elaborate design of time delays, and thus the control 
	of the near-field beam split is achievable.
	A simple analogy of the near-field 
	controllable beam split effect is the dispersion of white light with a 
	large bandwidth caused by a prism, thus in this paper, this effect is also 
	termed as ``\textbf{near-field rainbow}".
	This near-field rainbow effect allows us to generate multiple beams 
	focusing on multiple locations simultaneously by only one radio-frequency  
	(RF) chain, 
	which is not achievable for existing schemes.
	
%	Therefore, 
%	the coverage ranges of the spatial angle and distance over the entire 
%	bandwidth are controllable.
	
	\item Then, based on the mechanism of the near-field rainbow, a wideband 
	near-field beam training scheme is proposed to realize fast near-field CSI 
	acquisition. 
	In the proposed scheme, multiple beams focusing on multiple angles in a 
	given distance range are generated by time-delay circuits in each time 
	slot. Next, for different time slots, different distance ranges are 
	measured by fine-tuning the time-delay parameters. 
	By this means, the optimal spatial angle can be searched in a 
	frequency-division manner, while the optimal distance is obtained in a 
	time-division manner. Unlike the exhaustive near-field beam training scheme 
	that searches only 
	one location in each time slot, the proposed scheme is able to search 
	multiple 
	locations in each time slot, thus the beam training overhead can be    
	significantly reduced.}
	\item Finally, we provide simulation results to demonstrate the mechanism 
	of the near-field rainbow and verify the advantages of the proposed 
    beam training scheme. We demonstrate that our scheme is able 
	to achieve a satisfactory average rate performance at a significantly 
	reduced training overhead. 
	We also show that our beam training scheme outperforms the existing 
	far-field schemes in the near-field range.
	Additionally, our scheme also works well in the far-field range, since the proposed near-field beam training scheme is able to automatically decay to the far-field beam training in the far-field scenarios.
\end{itemize}

	\subsection{Organization and Notation}
	\emph{Organization}: The rest of this paper is organized as below. In 
	Section \ref{sec:2}, the system model is introduced. We first introduce the 
	near-field beam split effect in wideband XL-MIMO systems and then discuss 
	how to mitigate it with a time-delay beamformer. 
	Then in Section \ref{sec:3}, we prove 
	the mechanism of the near-field controllable beam split. In Section 
	\ref{sec:4}, the exhaustive near-field beam training is defined and the 
	near-field rainbow based beam training is proposed. Section \ref{sec:5} 
	provides the 
	simulation results. Finally, conclusions are drawn in Section \ref{sec:6}.
	
	\emph{Notation}: Lower-case and upper-case boldface letters represent vectors and matrices, respectively; 
%	$[\mb{X}]_{p,q}$ denotes the $(p, q)$-th entry of the matrix $\mb{X}$; 
	$[\mb{x}]_n$  denotes the $n$-th element of the vector $\mb{x}$;
	$(\cdot)^*$, $(\cdot)^T$, and $(\cdot)^H$ denote the conjugate, transpose, and conjugate transpose, respectively; $|\cdot|$ and $\text{Tr}(\cdot)$ denote the absolute and trace operator; 
	$\lceil x \rceil$ denotes the nearest integer greater than or equal to $x$;
	$\mathcal{CN}(\mu, \Sigma)$ and $\mathcal{U}(a,b)$
	denote the Gaussian distribution with mean $\mu$ and covariance
	$\Sigma$, and the uniform distribution between $a$ and $b$, respectively. 
%\clearpage
	\section{System Model} \label{sec:2}
In this section, we first introduce the near-field channel model and the 
near-field beam split effect in wideband XL-MIMO, then we discuss how to 
mitigate it by time-delay circuits.
\subsection{Near-Field Wideband Channel Model}
		\begin{figure}
	\centering
	%		\vspace*{-0.5em}
	\includegraphics[width=3.5in]{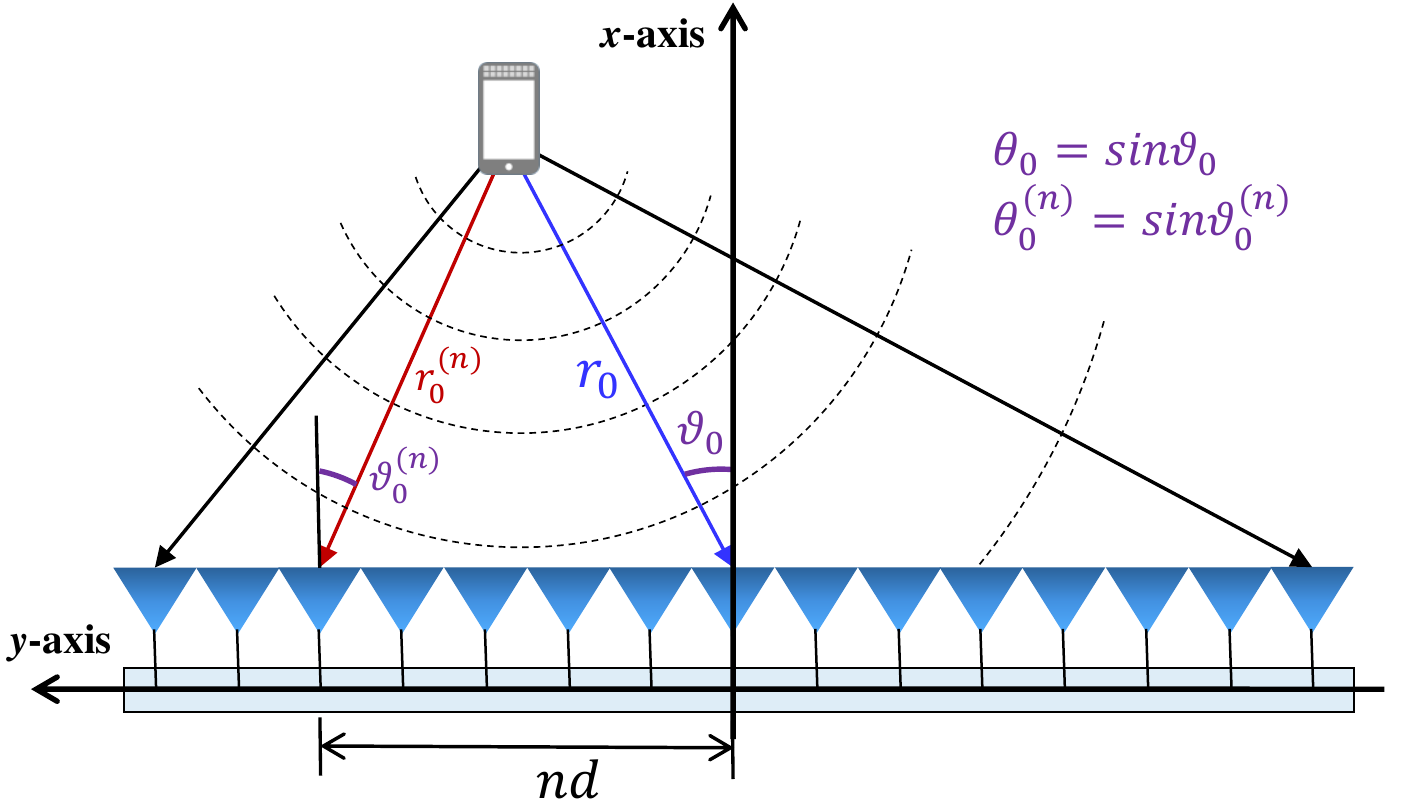}
	\caption{  The near-field spherical wavefront model.
	} 
	\label{img:channel}
	\vspace*{-1em}
\end{figure}
	In this paper, a wideband XL-MIMO system is considered. We assume the base 
	station (BS) is equipped with a uniform linear array (ULA) to serve a 
	single omnidirectional antenna user using orthogonal frequency division 
	multiplexing (OFDM) with $M$ subcarriers. 
	For expression simplicity, we assume the number of BS antennas is $N_t = 2 
	N + 1$.
	We denote $B$, $c$, $f_c$, 
	$\lambda_c = \frac{c}{f_c}$ as the bandwidth, the light speed, the central 
	carrier frequency, and the central wavelength, respectively. The antenna 
	spacing is 
	presented as $d = \frac{\lambda_c}{2}$. 
	
	{\color{black}Due to the severe path loss incurred by the scatters, mmWave 
	and THz 
	communications heavily rely on the line-of-sight (LoS) path 	
	\cite{THzsurvey_Elayan2020}. 
	Therefore, we mainly focus on the near-field LoS channel, 
	while the discussions in this paper can be straightforwardly extended to 
	the 
	non-LoS (NLoS) scenarios. 
	As shown in Fig. \ref{img:channel}, the user is located at 
	$(r_0\cos\vartheta_0, r_0\sin\vartheta_0)$ with $\theta_0 = 
	\sin\vartheta_0$.
	Taking into account the near-field 
	spherical wave characteristic \cite{NearLoS_Zhou2015}, the channel between 
	the $n$-th BS antenna and the user at the $m$-th subcarrier with $n = 
	-N,\cdots, 
	0,\cdots, N$ and $m = 
	1,2,\cdots, M$ can be represented as 
	\begin{align}\label{eq:channel0}
[\mb{h}_m]_n = \beta_m^{(n)}e^{-j{k_m r_0^{(n)}}},
	\end{align}}
	where 	$k_m = \frac{2\pi f_m}{c}$ 
	denotes the wavenumber at subcarrier frequency $f_m = f_c + 
	\frac{B}{M}\left(m - 1 - \frac{M - 1}{2}\right)$ and 
	$r^{(n)}_0$ represents the 
	distance between the user and the $n$-th BS antenna.
	As the coordinate of the $n$-th BS antenna is $(0, nd)$, 
	$r_{0}^{(n)}$ can be derived from the geometry as 
	$r_{0}^{(n)} = \sqrt{ (r_0\cos\vartheta_0)^2 + (r_0\sin\vartheta_0 - 
	nd)^2} = \sqrt{r_0^2 + n^2d^2 - 2r_l\theta_0nd}$.
	{\color{black}The path gain $\beta^{(n)}_m$ can be modeled as 
	\cite{RISNF_Tang2021}
	\begin{align}\label{eq:fspl}
	\beta^{(n)}_m = \sqrt{G_tF\left(\vartheta_0^{(n)}\right)} 
	\frac{\lambda_m}{4\pi r_0^{(n)}},
	\end{align} 
where wavelength $\lambda_m = \frac{f_m}{c}$ and $\vartheta^{(n)}_0$ denotes 
the angle between the user and the $n$-th BS antenna.
Moreover,  $G_t$ 
is the antenna gain and $F\left(\vartheta\right)$ denotes normalized power
radiation pattern, satisfying $\int_{\vartheta = 0}^{\pi/2} G_t 
F(\vartheta)\sin\vartheta d\vartheta = 1$. An example of normalized power 
radiation pattern is 
$F(\vartheta) = \cos^3\vartheta, \vartheta \in \left[-\frac{\pi}{2}, 
\frac{\pi}{2}\right]$ \cite{RISNF_Tang2021}.
Generally, the distance  $r_0$ between the user and BS is larger than the array 
aperture $D = N_t d$. For example, for a 256-element ULA working at 30 GHz,  
$r_0$ is very likely to be larger than $D = 256 \times 0.5\times10^{-3} = 1.28$ 
meters. With $r_0 > D$, we can assume $\beta^{(-N)}_m \approx\cdots\approx 
\beta^{(N)}_m \approx 
\beta_m =  \sqrt{G_tF\left(\vartheta_0\right)} 
\frac{\lambda_m}{4\pi r_0}$ 
based on the Fresnel approximation \cite{fresnel_Sherman1962}.
	As a result, the near-field LoS channel $\mb{h}_m\in\mathbb{C}^{N_t \times 
	1 }$ can be represented as 
	\begin{align}\label{eq:channel}
\mb{h}_m &= \beta_m\left[e^{-j{k_m r_0^{(-N)}}},\cdots, e^{-j{k_m r_0^{(N)}}} 
\right]^T  = \sqrt{N_t}\beta_m\mb{a}_m(\theta_0, r_0),
\end{align}
	where $\mb{a}_m(\theta_0, r_0)$ is the near-field array response vector.
}			\begin{figure}
	\centering
	%		\vspace*{-0.5em}
	\includegraphics[width=3in]{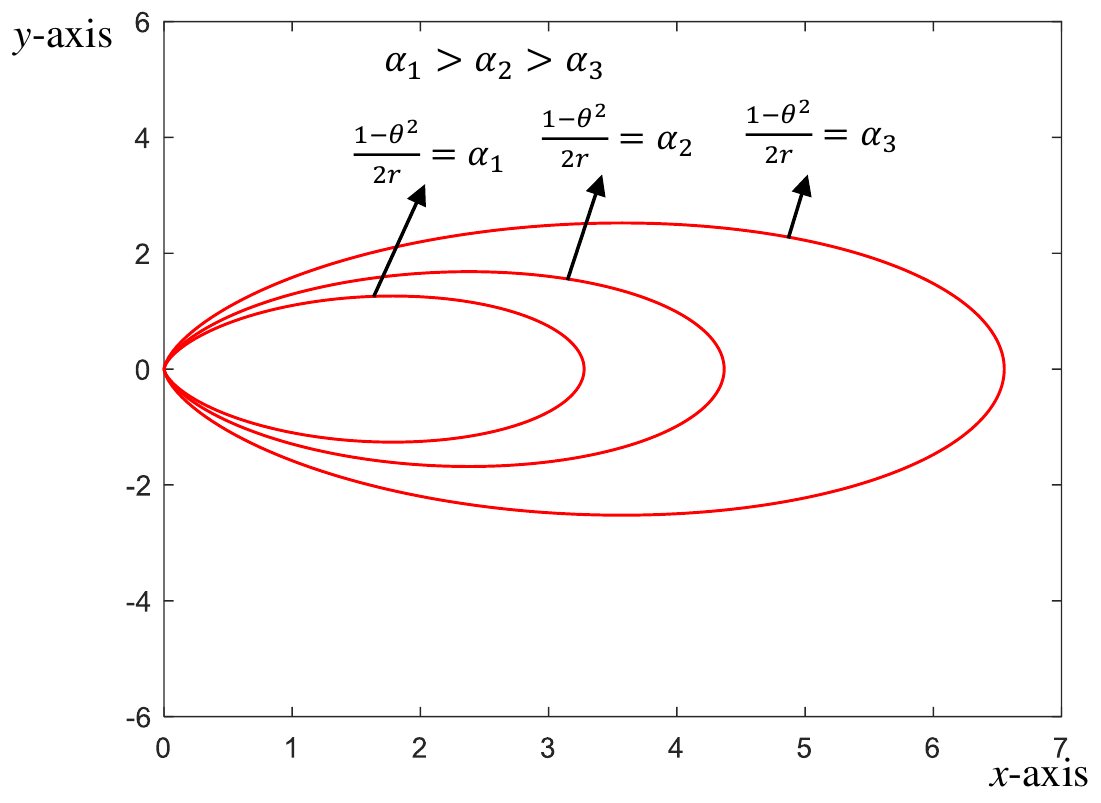}
	\caption{  The schematic diagrams of different distance rings $\alpha = 
	\frac{1 - 
	\theta^2}{2r}$.
	} 
	\label{img:curve}
	\vspace*{-1em}
\end{figure}

 Notice that since the near-field spherical wave characteristic is considered, 
 the array response vector $\mb{a}_m(\theta_0,r_0)$ is significantly different 
 from the classical far-field array response vector \cite{FundWC_Tse2015}, 
 where the latter is based on the planar wave assumption and ignores the 
 influence of 
 distance $r_0$. {\color{black}The radius of the near-field region is 
 determined 
 by the 
 Rayleigh 
 distance $\text{RD} = \frac{2D^2}{\lambda_c}$ \cite{fresnel_Selvan2017}. For a 
 1-meter diameter array 
 operating at 30 GHz, its Rayleigh distance reaches up to 200 meters. 
}Therefore, in 
 such a mmWave XL-MIMO system, the near-field spherical wave characteristic 
 becomes 
 essential and the distance information cannot be ignored.
 
However, since the distance $r_{0}^{(n)}$ is a complicated radical function 
with 
respect to the antenna index $n$, it is difficult to analyze the property 
of the near-field spherical wavefront directly from (\ref{eq:channel}). To deal 
		with this problem, the Fresnel approximation \cite{fresnel_Sherman1962} 
		can be adopted to approximate $r_{0}^{(n)}$ as 
	\begin{align}\label{eq:fresnel}
	r_{0}^{(n)} \overset{(a)}{\approx} r_0 - n d \theta_0 + n^2 d^2\frac{ 1 - 
	\theta_0^2}{2r_0},
	\end{align} 
	where $(a)$ is based on the second-order Taylor expansion $\sqrt{1 + x} 
	\approx 1 + \frac{1}{2}x - \frac{1}{8}x^2$. For expression simplicity, we 
	denote $\alpha_0 = \frac{1 - \theta_0^2}{2r_0}$ and $\alpha = \frac{1 - 
	\theta^2}{2r}$. As shown in 
	Fig. \ref{img:curve}, the curve $\alpha = \frac{1 - \theta^2}{2r}$ 
	corresponds to a ring in the physical space, and thus the curve $\alpha = 
	\frac{1 - 
	\theta^2}{2r}$  is termed as
	\emph{distance ring} $\alpha$. 
	Then, according to (\ref{eq:fresnel}), the $n$-th element of 
	$\mb{a}_m(\theta_0, r_0)$ 
	can be approximated as 
	\begin{align}
	[\mb{a}_m(\theta_0, r_0)]_n &\approx \frac{1}{\sqrt{N_t}}e^{-jk_m(r_0 -
	n d \theta_0 +  n^2 d^2 \alpha_0)} 
	= e^{-jk_mr_0} 
	[\mb{b}_m( \theta_0,\alpha_0)]_n, \label{eq:Fresnel}
	\end{align}
	where $[\mb{b}_m(\theta_0, \alpha_0)]_n 
	= \frac{1}{\sqrt{N}} e^{ j k_m (nd \theta_0 - n^2 d^2 
	\alpha_0)}$ denotes the $n$-th element of vector
	$\mb{b}_m(\theta_0, \alpha_0)$. Since the residue phase 
	$e^{-jk_mr_0}$ 
	in (\ref{eq:Fresnel}) has no relationship with the antenna index $n$, we 
	only need to focus on the vector $\mb{b}_m(\theta_0, \alpha_0)$.

\subsection{Near-Field Beam Split}
	At mmWave or THz band, \emph{frequency-independent} PS-based beamformer 
 is widely considered  to serve the user by generating a 
	focused beam aligned with 
	location $(r_0, \theta_0)$ in the polar coordinate. 
	Generally, the beamfocusing vector $\mb{w}_{\text{PS}}$ generated by  
	{frequency-independent} PSs is usually set as 
	$\mb{w}_{\text{PS}} = \mb{b}_c^*(\theta_0, \alpha_0)$ to 
	focus the beam energy on the location $(r_0, \theta_0)$ 
	\cite{THzbeam_Headland2018}, where 
	$\mb{b}_c(\theta_0, \alpha_0)$ represents the array response vector 
	(\ref{eq:Fresnel}) at 
	frequency $f_c$.
	
		\begin{figure}
		\centering
		%		\vspace*{-0.5em}
		\includegraphics[width=5in]{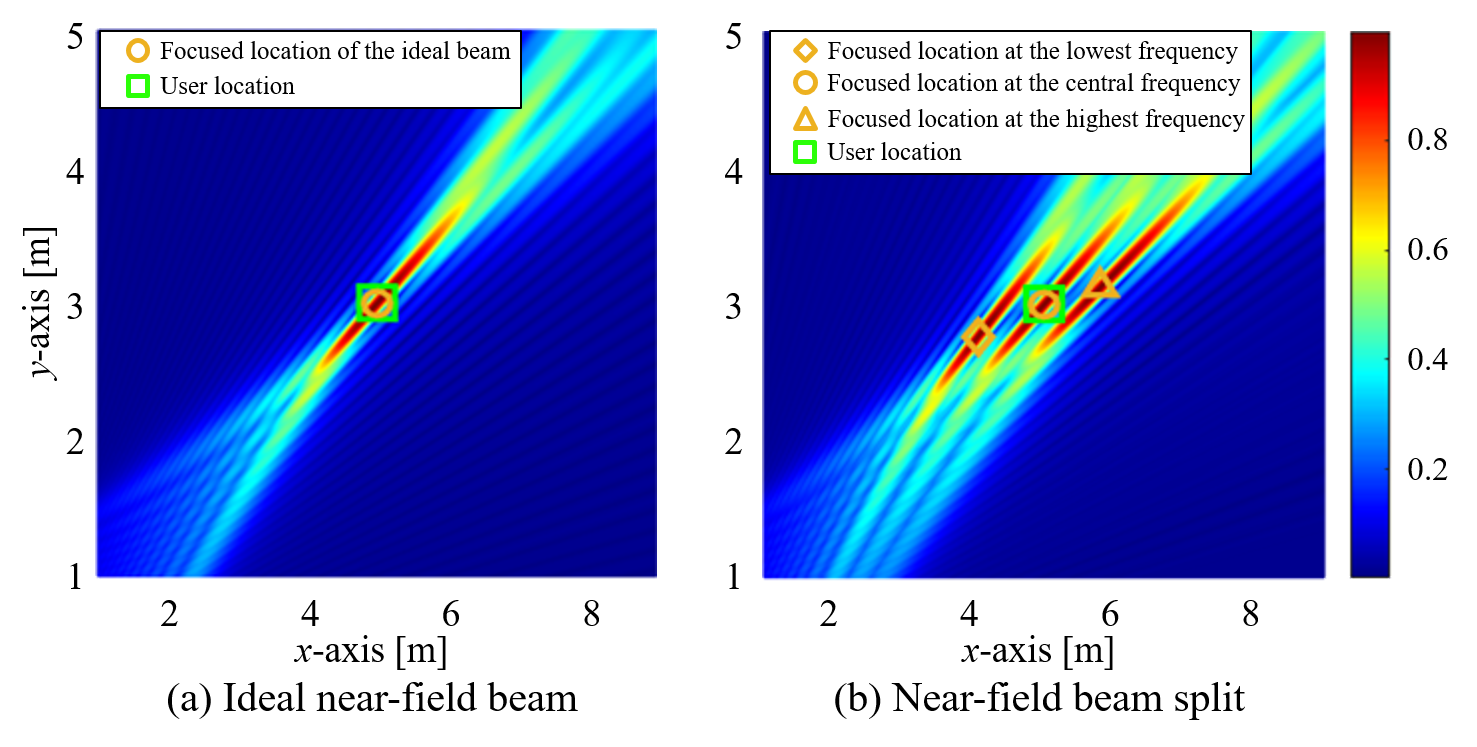}
		\caption{ {\color{black} The near-field beam split effect.
		} }
		\label{img:NFBS}
		\vspace*{-1em}
	\end{figure}

	However, the array response vector (\ref{eq:Fresnel}) of the wideband 
	channel is \emph{frequency-dependent}, which mismatches the 
	\emph{frequency-independent} beamfocusing vector $\mb{w}_{\text{PS}} = 
	\mb{b}_c^*(\theta_0, \alpha_0)$. This mismatch leads to the fact that,  
	the beams generated by $\mb{w}_{\text{PS}}$ at different 
	frequencies will be focused on different locations, which is detailed in 
	the following \textbf{Lemma 1}. 
 \begin{thm} \label{lemma1}
 	For near-field wideband communications, the beam at frequency $f_m$ 
 	generated by $\mb{w}_{\text{PS}} = 
 	\mb{b}_c^*(\theta_0, \alpha_0)$ will be focused on the location $(\theta_m, 
 	r_m)$, satisfying  
	\begin{align}
\theta_m &= (k_c/k_m) \theta_0 = (f_c/f_m) \theta_0 = \theta_0 / \eta_m, 
\label{eq:theta_phase}\\
\alpha_m &= (k_c/k_m) \alpha_0 = (f_c/f_m) \alpha_0 = \alpha_0 
\label{eq:alpha_phase}/ \eta_m,
\end{align} 
	where we define $\alpha_m = \frac{1 - 
		\theta_m^2}{2r_m}$, $\eta_m = f_m / f_c$, and $k_c = \frac{2\pi 
		f_c}{c}$.
\end{thm}	
\emph{Proof}: At frequency $f_m$, 
	the array gain achieved by $\mb{w}_{\text{PS}}$ on an arbitrary user 
	location $(r, \theta)$ with $\alpha = \frac{1 - \theta^2}{2r}$ is 
%	we evaluate the array gain on an arbitrary user location $(r, \theta)$ with $\alpha = \frac{1 - \theta^2}{2r}$, which is given by 
	\begin{align}\label{eq:G}
	|\mb{w}_{\text{PS}}^T\mb{b}_m(\theta, \alpha)| &= \frac{1}{N_t} 
	\left|\sum_{n 
	= -N}^{N }  e^{ jn d (k_m \theta - k_c \theta_0) - jn^2 d^2 
	(k_m \alpha - k_c \alpha_0)} \right| \notag\\
	&= G(k_m\theta - k_c \theta_0, k_m \alpha - k_c \alpha_0), 
	\end{align}
	where we define $G(x, y) = \frac{1}{N_t} \left|\sum_n  e^{ jn d x - 
	jn^2 d^2 y} \right|$. 
	Apparently, the beam at frequency $f_m$ is focused on the location $(r_m, 
	\theta_m)$ corresponding to 
	the maximum array gain $G(k_m\theta_m - k_c \theta_0, k_m \alpha_m - k_c 
	\alpha_0)$, i.e., 
	\begin{align} \label{eq:Gmax}
	(r_m, \theta_m) = \arg\max_{r, \theta} G(k_m\theta - k_c \theta_0, k_m 
	\alpha - k_c \alpha_0),
	\end{align}
	where $\alpha_m = \frac{1 - \theta_m^2}{2r_m}$.
	Since $G(0, 0) = 1$ and $|G(x,y)|\le1$, it is obvious that $(x, y) = (0,0)$ 
	is an optimal solution to maximize the function $G(x ,y)$. Therefore, we 
	have $k_m\theta_m - k_c \theta_0 = 0$ and $k_m \alpha_m - k_c \alpha_0 = 
	0$, which leads to the results of (\ref{eq:theta_phase}) and 
	(\ref{eq:alpha_phase}).
$\hfill\blacksquare$

According to \textbf{Lemma 1}, the beam at $f_m$ is focused on the location 
$(\theta_m, r_m) = (\theta_m, \frac{1 - \theta_m^2}{2\alpha_m}) = 
(\frac{\theta_0}{\eta_m}, r_0 \frac{\eta_m - 
	\frac{1}{\eta_m}\theta_0^2}{1 - \theta_0^2} )$. As shown in Fig. 
	\ref{img:NFBS}, since the desired user is located at $(r_0, \theta_0) \neq 
	(r_m, \theta_m)$, the beam at frequency $f_m$ cannot be aligned with the 
	desired 
	user location. {\color{black}
	This misalignment is termed as the ``near-field beam split" effect 
	\cite{NFBF_Cui2021}, which will result in a severe array gain loss when 
	$f_m$ is far away from $f_c$ in 	wideband XL-MIMO systems. 
	For example, if the carrier frequency and bandwidth are $30$ GHz and $1$ 
	GHz, and the BS is equipped with a 256-element ULA, then around 50\% 
	subcarriers will suffer from more than 50\% array gain loss 	
	\cite{NFBF_Cui2021}.}
	Therefore, the near-field beam split effect should be elaborately 
	addressed, especially when the bandwidth is very large.
	
\subsection{Time-Delay Based Beamformer}\label{sec:2-2}
To mitigate the near-field beam split effect, a method is to utilize 
TD based beamformer rather than PS based beamformer to generate 
frequency-dependent beams to match the frequency-dependent 
channels. Notice that the TD based beamsteering has been studied in the 
far-field wideband systems 	\cite{TTDhardware_Hashemi2008, TTD_Gha2019, 
TTD_Lin2017, DPP_Tan2019, DPP_Dang2021, TTDBT_Yan2019, TTDBT_Bol2021, 
THzBT_Tan2021}, while in this study, we consider utilizing it to 
overcome and control the near-field beam split effect. 

	\begin{figure}
	\centering
	%		\vspace*{-0.5em}
	\includegraphics[width=3.5in]{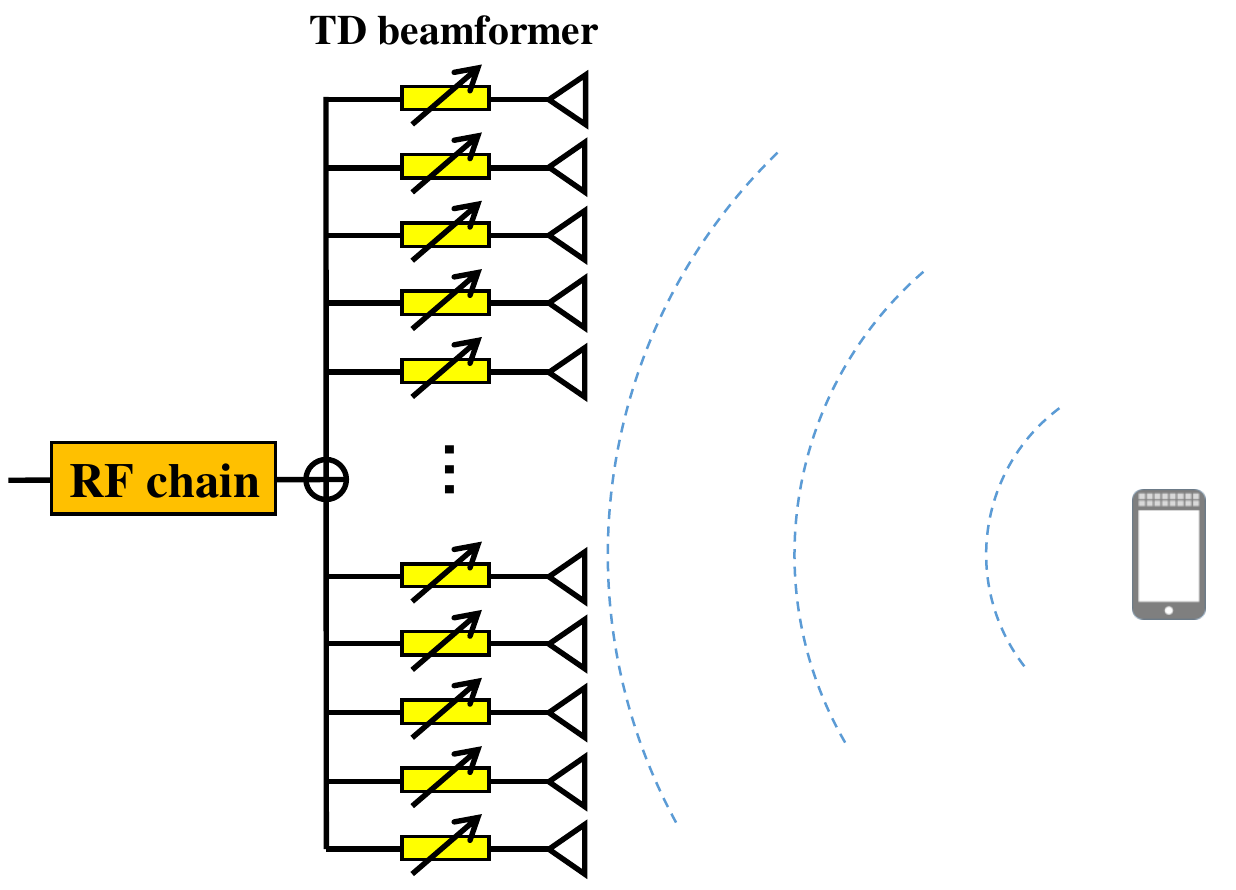}
	\caption{  The TD beamformer architecture.
	} 
	\label{img:TTD}
	\vspace*{-1em}
\end{figure}

As shown in Fig. \ref{img:TTD}, we assume the BS is equipped with an 
$N_t$-element TD beamformer\footnote{\color{black}The main purpose of this 
paper 
is to 
	derive that time-delay circuits are able to control the near-field beam 
	split 
	effect. Therefore, we directly utilize the time-delay beamformer as an 
	example, 
	while the discussion on some other low-power consumption architectures, 
	such as 
	delay-phase precoding,  is left for future works.}. Then the frequency 
	response at frequency $f_m$ 
of 
the $n$-th antenna is $[\mb{w}_m]_n = \frac{1}{\sqrt{N_t}}e^{-j2\pi f_m 
\tau^{(n)'}}$, where  $\mb{w}_m$ denotes the frequency-dependent beamforming 
vector, and $\tau^{(n)'}$ denotes the adjustable time-delay parameter of the 
$n$-th time-delay circuit. For expression simplicity, we denote $r^{(n)'} = 
c\tau^{(n)'}$ as the adjustable distance, and then $[\mb{w}_m]_n$ becomes  
$\frac{1}{\sqrt{N_t}}e^{-jk_m r^{(n)'}}$. Notice that $[\mb{w}_m]_n$ has a 
similar 
form with the array response vector $[\mb{b}_m(\theta_0, r_0)]_n = 
\frac{1}{\sqrt{N_t}}e^{jk_m (nd\theta_0 - n^2d^2\alpha_0)}$. Thus, the 
$n$-th adjustable distance $r^{(n)'}$ can be set as $r^{(n)'} = 
nd\theta' -  n^2d^2\alpha'$, while $\theta'$ and $\alpha'$ are 
adjustable delay parameters of a TD beamformer. Then, the corresponding 
beamfocusing vector at $f_m$ can be presented as
\begin{align}\label{eq:TTD}
&[\mb{w}_m(\theta', \alpha')]_n  =\frac{1}{\sqrt{N_t}}  e^{-jk_m (nd\theta' - 
n^2 
d^2 \alpha' )}. 
\end{align} 
In this case, at frequency $f_m$, the array gain on an arbitrary location $(r, 
\theta)$ with $\alpha = \frac{1 - \theta^2}{2r}$ is given by
\begin{align}\label{eq:TTDG}
	|\mb{w}_m(\theta', \alpha')^T\mb{b}_m(\theta, \alpha)| &= 
	\frac{1}{N_t} 
	\left|\sum_{n = -N}^{N}  e^{ jn d (k_m \theta - k_m \theta') - 
	jn^2 d^2 (k_m \alpha - k_m \alpha')} \right| \notag\\
&= G(k_m(\theta - \theta'), k_m (\alpha - \alpha')). 
\end{align} 
Obviously, the beam at $f_m$ is focused on $(r_m, \theta_m) = \arg\max_{r, \theta} G(k_m(\theta - \theta'), k_m (\alpha - \alpha'))$.
Similar to the derivation of the optimal solution of (\ref{eq:Gmax}),  
$G(k_m(\theta - \theta'), k_m (\alpha - \alpha'))$ approaches its maximum value 
when $k_m(\theta - \theta') = 0$ and  $k_m (\alpha - \alpha') = 0$. 
Therefore, the beam at frequency $f_m$ is focused on $\theta_m = \theta'$ and $r_m = r' = \frac{1 - \theta'^2}{2\alpha'}$, which is not relevant to $f_m$.
Then, once the LoS path information $(r_0, \theta_0)$ is available at 
the BS, and the delay parameters are set as $\theta' = \theta_0$ and $\alpha' = 
\frac{1 - \theta_0^2}{2r_0}$, then the beams across the entire bandwidth are 
able to be focused on the location $\theta_m = \theta' = \theta_0$ and $r_m = 
\frac{1 - \theta'^2}{2\alpha'} =  r_0$. As a consequence, the near-field beam 
split effect can be mitigated by  the TD beamformer.

On the other hand, efficient wideband beamfocusing requires that the LoS path 
information $(r_0, \theta_0)$ is available at the BS side. 
To realize this requirement, in the following discussions, we prove that a TD  
beamformer can not only mitigate the near-field beam split effect, but also 
flexibly control its degree.
Then, we will further utilize this property to achieve efficient near-field beam training to obtain $(r_0, \theta_0)$.
 
	\section{Mechanism of Near-Field Controllable Beam Split} \label{sec:3}
	{\color{black}In \cite{TTDBT_Yan2019, THzBT_Tan2021, TTDBT_Bol2021}, the 
	researchers 
	proved the mechanism of far-field controllable beam split, i.e., by 
	elaborately designing the time-delay parameters, the beams over the 
	entire bandwidth are able to cover a desired angular range in the 
	far-field. 
	Taking advantage of this mechanism, multiple beams aligned with multiple 
	angles can be generated by only one RF chain to acquire the far-field CSI 
	rapidly.
	Similarly, to obtain the near-field CSI, we surprisingly find that 
	time-delay circuits can  not only control the angular coverage range but 
	also control the distance coverage range of the beams, i.e., the near-field 
	controllable beam split is achievable. In this section, we will prove the 
	mechanism of this effect, and then in the next section 
	utilize this mechanism to achieve fast near-field beam 
	training. For readers better understand our ideas, we first prove the 
	far-field controllable beam split and then extend it to 
	the near-field scenarios.}
\subsection{Far-Field Controllable Beam Split} \label{sec:3-1}
For the far-field scenarios, the distances $r$ and $r'$ are assumed to be 
larger than the Rayleigh distance $\text{RD}$, so that the spherical wavefront 
can be approximated as a planar wavefront. In this case, all of the 
distance-related 
parameters $\alpha = \frac{1 - \theta^2}{2r}$ and $\alpha' = \frac{1 - 
\theta'^2}{2r'}$ reliably approach 0.
Then, the array response vector becomes $[\mb{b}_m(\theta, \alpha)]_n = 
[\mb{b}_m(\theta, 0)]_n = \frac{1}{\sqrt{N_t}} e^{jk_m n d \theta}$, and 
the beamfocusing vector realized by a TD beamformer becomes 
$[\mb{w}_m(\theta', 0)]_n = \frac{1}{\sqrt{N_t}} e^{-jk_m n d \theta'}$. 
Therefore, the array gain in (\ref{eq:TTDG}) can be simplified as 
% and the array gain function in (\ref{eq:TTDG}) can be simplified as $G(k_m(\theta - \theta'), 0)$.
\begin{align}\label{eq:FarG}
G\left(k_m(\theta - \theta'), 0\right) = \left|\frac{\sin\left( 
\frac{N_t}{2}dk_m(\theta - \theta')\right)}{N_t\sin\left( 
\frac{1}{2}dk_m(\theta - \theta')\right)}\right|,
\end{align}
where $G(x, 0) = \left|\frac{\sin\left( \frac{N_t}{2}dx\right)}{N_t\sin\left( 
\frac{1}{2}dx\right)}\right|$. 
It has been previously proved that the beams over the entire bandwidth 
generated by $\mb{w}_m(\theta', 0)$ are focused on the spatial angle $\theta_m 
= \theta'$, with $m = 1,2,\cdots, M$.
However, notice that the spatial angle $\theta_m$ is corresponding to an actual 
physical angle $\vartheta_m = \arcsin\theta_m$ as shown in Fig. 
\ref{img:channel}, which implies the value range of $\theta_m$ is restrained by 
$\theta_m \in [-1, 1]$. By contrast, $\theta'$ is an adjustable parameter of a 
TD beamformer. A question naturally arises that, if the adjustable parameter is 
set as $\theta' \notin [-1, 
1]$, then what it has to do with the actual spatial angle $\theta_m$? Actually, 
the answer to this question is exactly the mechanism of the far-field 
controllable beam split.

For expression simplicity, parameter $\theta' \notin [-1, 1]$ is termed as the 
abnormal value, while parameter $\theta' \in [-1, 1]$ is termed as the 
actual value. Restrained by the value range of $\theta_m$, if 
$\theta'$ is an 
abnormal value, it is obvious that $\theta_m \neq \theta'$. 
To acquire the relationship between $\theta_m$ and $\theta'$, we 
observe that the far-field array gain $G(x, 0)$ is a periodic function against 
$x$ with a period $\frac{2\pi}{d}$. For any integer $p \in 
\mathbb{Z}$, we have
\begin{align}\label{eq:far_period}
G\left(x - \frac{2p\pi}{d}, 0\right) &=  \left| \frac{\sin\left( \frac{N}{2}dx - Np\pi \right)}{N\sin\left( \frac{1}{2}dx -
	p\pi\right)}\right| =\left|\frac{\sin\left( \frac{N}{2}dx 
	\right)}{N\sin\left( \frac{1}{2}dx\right)} \right|  =
	 G(x, 0).
\end{align}
In Section \ref{sec:2}, we have indicated that the optimal solution 
$x^{\text{opt}}$ to maximize $G(x, 0)$ is $(x^{\text{opt}}, 0) = (0,0)$. 
However, according to (\ref{eq:far_period}), we find that  $(0,0)$ is just one 
of the 
optimal solutions for maximizing $G(x, 0)$. 
The periodicity of $G(x, 0)$ implies that the optimal solutions should 
satisfy $(x^{\text{opt}}, 0) = (\frac{2p\pi}{d},0)$, $p \in \mathbb{Z}$. 
As a consequence, by solving $\theta_m = \arg\max_{\theta}G(k_m(\theta - 
\theta'), 0)$, the focused spatial angle $\theta_m$ at frequency $f_m$ is 
derived as 
\begin{align} \label{eq:thetaf}
\theta_m = \theta' + \frac{2 p \pi}{dk_m} = \theta' + \frac{2 p}{\eta_m}, 
\end{align} 
where the function of $p \in \mathbb{Z}$ is to ensure $\theta_m \in 
[-1, 1]$ becomes an actual spatial angle. 
From (\ref{eq:thetaf}), the mechanism of the far-field controllable beam split 
is acquired, which has the following features.

\textbf{Feature 1}: Since $p$ is an integer, if $ p \neq 0$, the spatial angle 
$\theta_m$ is related to the frequency $f_m$, which means beams at different 
frequencies will split towards different spatial angles. Therefore, despite the 
TD based beamsteering architecture being utilized, the far-field 
beam split effect can also be induced.

\textbf{Feature 2}: 
The specific value of $p \in \mathbb{Z}$ is determined by 
$\theta'$. If $\theta' \in [-1, 1]$ belongs to the actual value range of 
$\theta_m$, then $p$ can be zero and $\theta_m$ exactly equals to $\theta'$. 
That is to say, with $\theta' \in [-1, 1]$, the far-field beam split effect is 
eliminated. While if  $\theta' \notin [-1, 1]$ becomes an abnormal value, 
then to guarantee that $\theta_m$ is an actual spatial angle, it is obvious 
that $p$ cannot be zero. For instance, if $\theta'$ is set as $1.5$, at the 
central frequency $f_c$ with $\eta_c = \frac{f_c}{f_c} = 1$, the corresponding 
spatial angle  is 
$\theta_c = 1.5 + 2p/\eta_c = 1.5 + 2p$, $p\in \mathbb{Z}$. Only if $p = -1$ 
can $\theta_c = 
-0.5 \in [-1, 1]$ be an actual spatial angle. Therefore, with $\theta' \notin 
[-1, 1]$, $p$ is not zero and thus $\theta_m$ is a function of the frequency 
$f_m$, where the beam split effect is induced again.

\textbf{Feature 3}: Notice that the beam split pattern (\ref{eq:thetaf}) of a 
TD beamformer is different from (\ref{eq:theta_phase}) of a PS 
beamformer. For a PS beamformer, the degree of beam split is fixed and 
uncontrollable. However, for a TD beamformer, the degree of beam split is 
controllable by adjusting $\theta'$. For instance, considering the central 
frequency $f_c$, if $\theta'$ is set as $1.5$, then $p$ should be $-1$ and 
$\theta_c = 1.5 - 2 = -0.5$. On the other hand, if $\theta'$ is set as $3.5$, 
then $p$ should be $-2$ and $\theta_c = 3.5 - 4 = -0.5$. For the above two 
examples, although the spatial angles $\theta_c$ are the same, the integers $p$ 
are different, which indicates the degree of the beam split $\theta_m = \theta' 
+ 
2p/\eta_m$ is different. 
{\color{black}Therefore, by adjusting the delay parameter $\theta'$, the 
integer 
$p$ is 
indirectly controlled, then the degree of beam split is adjusted. This property 
is termed as the far-field controllable beam split.
Benefiting from this property, the adjustment on the angular coverage range of 
the beams is achieved, which can be utilized to realize fast far-field CSI 
acquisition \cite{TTDBT_Yan2019, THzBT_Tan2021, TTDBT_Bol2021}.
}

In the next sub-section, we will extend this conclusion to a more general near-field scenario.

\subsection{Near-Field Controllable Beam Split} \label{sec:3-2}
For the near-field scenario, the distances $r$ and $r'$ are lower than the Rayleigh distance. Then $\alpha$ and $\alpha'$ can not be zero, and the near-field spherical wave characteristics should be considered. 
%For this scenario, we prove that a time-delay array can not only control the 
%degree of beam split on the spatial angle $\theta_m$, but also control the 
%degree of beam split on the distance $r_m$ or $\alpha_m$. Therefore, this 
%phenomenon is referred to as the NF-CBS .
To prove the mechanism of the near-field controllable beam split, analog to the 
derivation in the far-field scenarios, we consider the abnormal values for 
the adjustable parameters $\theta'$ and $\alpha'$ in (\ref{eq:TTD}) 
simultaneously.

Specifically, it has been discussed before $\theta' \notin [-1, 1]$ is an 
abnormal value for the angle. As for the distance-related parameter $\alpha = 
\frac{1 - \theta^2}{2r}$, since $1 > \theta^2$ and $r > 0$, a realistic 
$\alpha$ should be larger than zero. On this 
condition, $\alpha' < 0$ is obvious the abnormal value for the distance ring. 
If 
$\theta' \in [-1, 1]$ and $\alpha' > 0$ are actual values, it was proved  in 
Section \ref{sec:2-2} that the near-field beam split effect is eliminated, 
i.e., beams are focused on 
$\theta_m = \theta'$ and $\alpha_m = \alpha'$ for all frequencies $f_m$. 
On the other hand, for the abnormal value $\theta' \notin [-1, 1]$ or $\alpha' 
< 0$, 
the following \textbf{Lemma 2} explains the behaviors of the beams 
generated by $\mb{w}_m(\theta', \alpha')$.
 \begin{thm} \label{lemma2}
 	If $\theta' \notin [-1, 1]$ or $\alpha' < 0$, the beam at frequency $f_m$ 
 	generated by 
 	$\mb{w}_m(\theta', \alpha')$ according to (\ref{eq:TTD}) will be focused on 
 	the location $(\theta_m, r_m)$, satisfying  
\begin{align} 
\theta_m &= \theta' + \frac{2 p \pi}{dk_m} = \theta' + \frac{2 
p}{\eta_m}, \label{eq:theta2} \\
\alpha_m &= \alpha' + \frac{2 q \pi}{d^2k_m} = \alpha' + \frac{2 
q}{d\eta_m},\label{eq:alpha2}
\end{align} 
where $r_m = \frac{1-\theta_m^2}{2\alpha_m}$ and $p,q \in \mathbb{Z}$ are 
integers to guarantee $\theta_m \in [-1, 1]$ and $\alpha_m > 0$.
 \end{thm}
 \emph{Proof}: Similar to the derivation of (\ref{eq:thetaf}), the periodicity 
 of the near-field array gain function $G(x, y)$ should be considered. It 
is proved in Appendix A that $G(x, y)$ is a periodic function against the 
vector variable $(x, y)$ with a period $(\frac{2\pi}{d}, \frac{2\pi}{d^2})$. 
Therefore, for any integers $p\in\mathbb{Z}$ and $q \in \mathbb{Z}$, 
$G(x, y)$ can be rewritten as 
\begin{align}\label{eq:Gnear}
G(x, y) =  G(x - \frac{2p\pi}{d}, y - \frac{2q\pi}{d^2}).
\end{align}

It has been indicated in Section \ref{sec:2} that the optimal solution 
$(x^{\text{opt}}, y^{\text{opt}})$ to maximize the array gain  $G(x, y)$ is 
$(x^{\text{opt}}, y^{\text{opt}}) = (0,0)$.
The periodicity in (\ref{eq:Gnear}) implies $(0,0)$ is just one of the optimal 
solutions for maximizing $G(x, 
y)$. Besides, the optimal solutions should satisfy 
$(x^{\text{opt}}, y^{\text{opt}}) = (\frac{2p\pi}{d},\frac{2q\pi}{d^2})$ with 
$p \in \mathbb{Z}$ and $q \in \mathbb{Z}$. Accordingly, to obtain the focused 
location of 
the beam at frequency $f_m$ by solving $(\theta_m, r_m) = \arg\max_{\theta, 
r}G(k_m(\theta 
- \theta'), k_m(\alpha - \alpha'))$, we have $k_m(\theta_m - \theta') = \frac{2 
p \pi}{d}$ and $k_m(\alpha_m - \alpha') = \frac{2 q \pi}{d^2}$. 
As a consequence, $\theta_m$ and $\alpha_m$ become $\theta' + \frac{2 
	p}{\eta_m}$ and $\alpha' + \frac{2 
	q}{d\eta_m}$, respectively, where the function of integers $p$ and $q$ is 
	to ensure that $\theta_m \in [-1, 1]$ and $\alpha_m > 0$ 
are an actual spatial angle and an actual distance ring.
$\hfill\blacksquare$

According to \textbf{Lemma 2}, the mechanism of the near-field controllable 
beam split is acquired, which has the following features.
 	\begin{figure}
	\centering
	%		\vspace*{-0.5em}
	\subfigure[]
	{\includegraphics[width=3in]{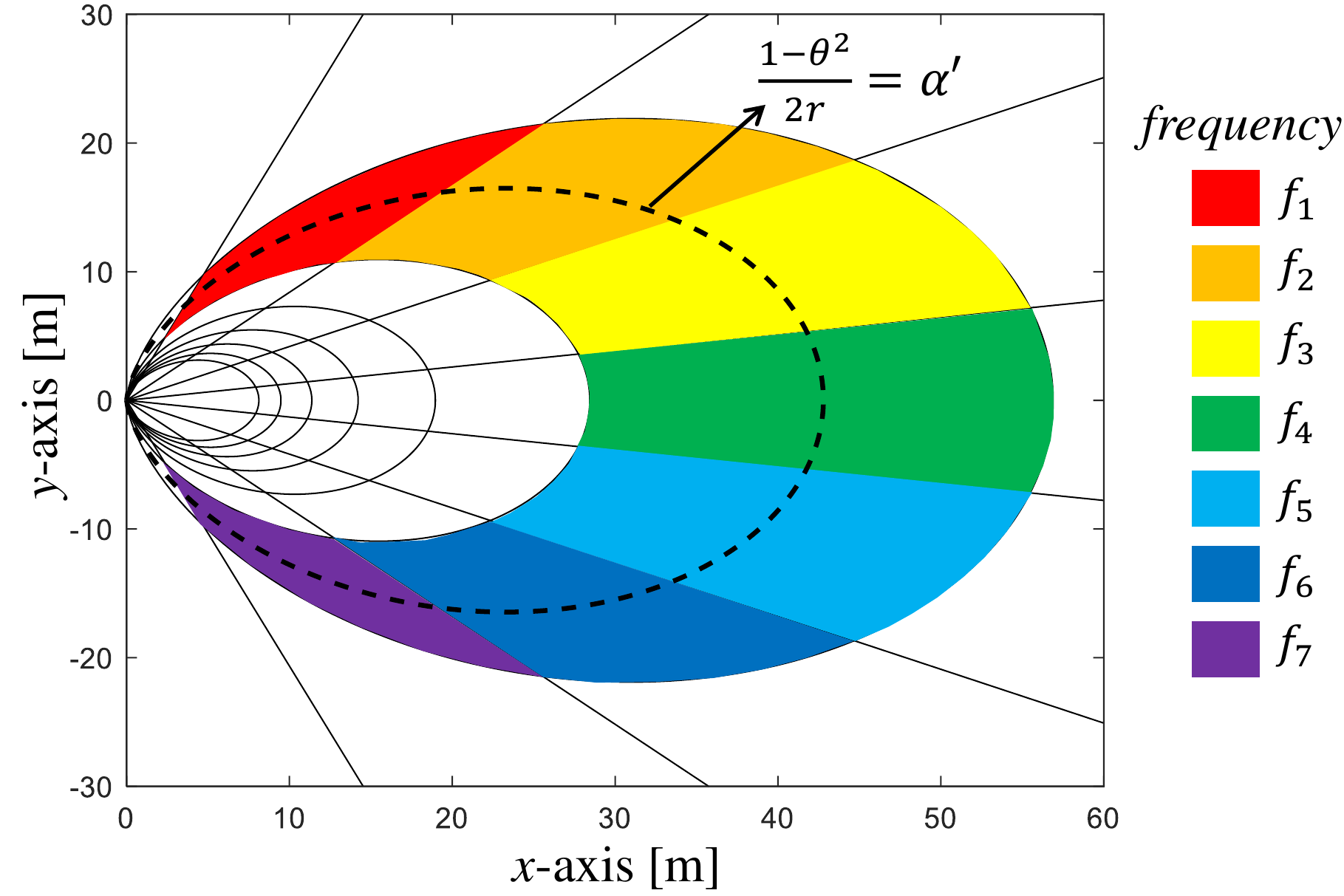}}
	\subfigure[]
	{\includegraphics[width=3in]{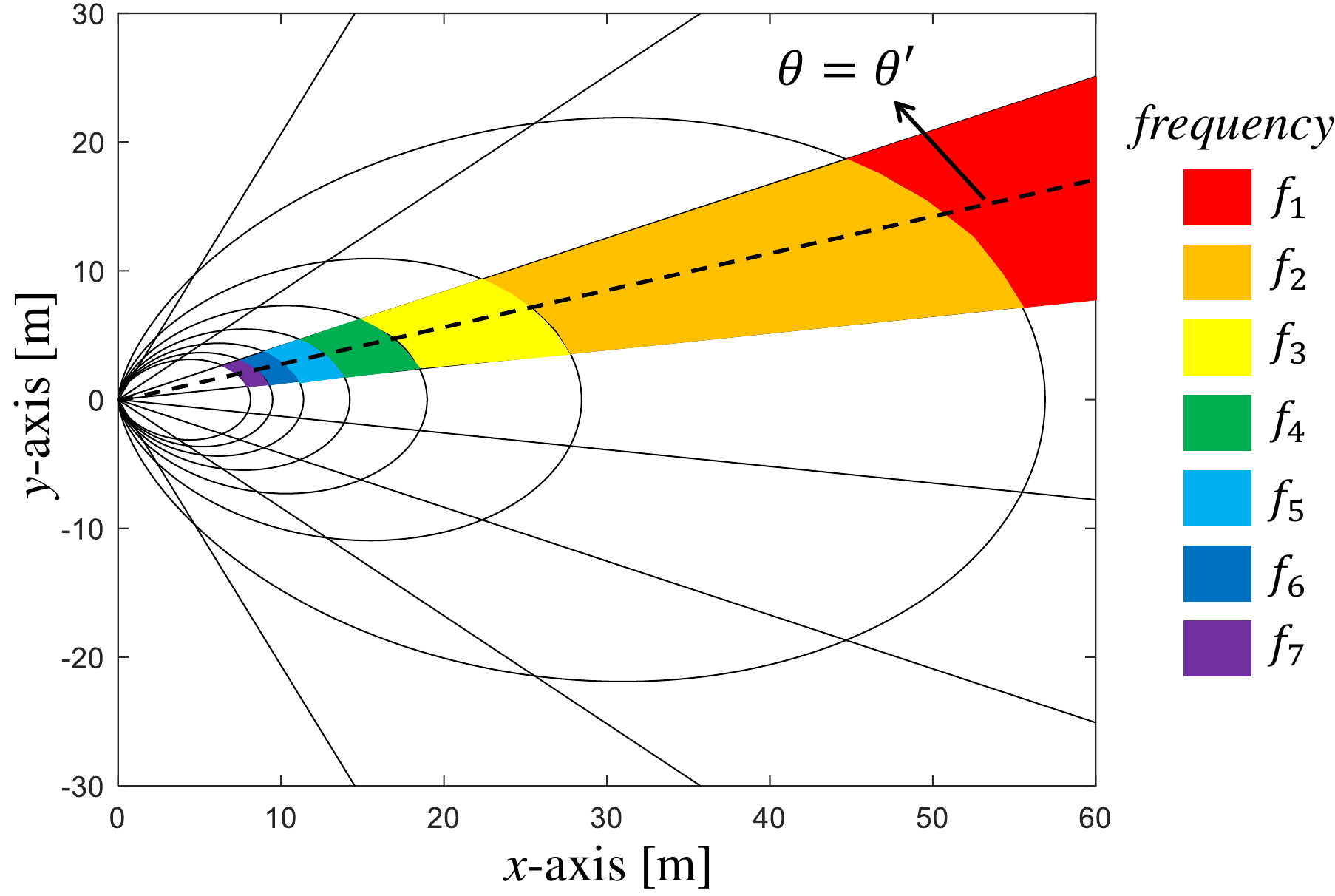}} 
	\caption{  \color{black}The schematic diagrams of the near-field rainbow on 
	(a) the  
	angle 
		dimension  and (b) the distance dimension.
	} 
	\label{img:rainbow}
	\vspace*{-1em}
\end{figure}

\textbf{Feature 1}: If $ p \neq 0$ or $q \neq 0$, then $\theta_m$ or $\alpha_m$ 
is related to the frequency $f_m$, which means beams at different frequencies 
will be focused on different locations. Therefore, despite the TD beamformer 
being utilized, the near-field beam split effect can also be induced.

\textbf{Feature 2}: As the function of the abnormal value of $\theta'$ has been 
discussed in the previous section, we mainly explain the impact of the abnormal 
value of $\alpha'$ on the integer $q$ here. 
If $\alpha' > 0$ is an actual value, then $q$ could be zero,  
and for all beams, we have $\alpha_m = \alpha'$. However, if $\alpha' < 0$ is 
an abnormal value, to guarantee 
$\alpha_m > 0$, the value of $q$ must be larger than 0, and thus the beams at 
different subcarriers will be focused on different distances $r_m = \frac{1 - 
\theta_m^2}{2\alpha_m}$. For instance, let the delay parameter $\alpha'$ 
be $-\frac{2}{d}$, then we have 
$\alpha_m = \frac{2}{d}(\frac{q}{\eta_m} - 1)$. To guarantee $\alpha_m \ge 0$ 
is an actual distance ring, the integer $q$ should be larger 
than $\eta_m$. 
Accordingly, when $f_m \le f_c$ and 
$\eta_m = \frac{f_m}{f_c} \le 1$, then the integer $q = 1$ is enough to make 
$\alpha_m > 0$
and the beam at $f_m$ is focused on the distance ring $\alpha_m = 
\frac{2}{d}(\frac{1}{\eta_m} - 1)$.
On the other hand, when $f_m > f_c$ and 
$\eta_m = \frac{f_m}{f_c} > 1$, then the integer $q = 2$ is enough to make 
$\alpha_m > 0$
and the beam at $f_m$ is focused on the distance ring $\alpha_m = 
\frac{2}{d}(\frac{2}{\eta_m} - 1)$.
As a consequence, the abnormal value of $\alpha'$ is able to introduce and 
control the beam split effect on the distance dimension.

\textbf{Feature 3}: 
The far-field scenario is a special case of the near-field controllable beam 
split by setting $\alpha_m = 
\alpha' = 0$.
The far-field scenario assumes the distance is very large, and it can only 
control the beam split effect on the angle $\theta_m$. By contrast, the 
near-field controllable beam split effect
can  adjust the degree of the beam split effect on the angle $\theta_m$ and 
distance $r_m$ jointly. For instance, one can set $\theta'$ as an abnormal 
value and  $\alpha'$ as an actual value, then we have $\theta_m = \theta' + 2p 
/ \eta_m$, $\alpha_m = \alpha'$ and $r_m = \frac{1 - \theta_m^2}{2\alpha_m} 
=\frac{1 - (\theta' + 2p/\eta_m)^2}{2\alpha'} $. Therefore, the beam at 
frequency $f_m$ is focused on a location satisfying $\left(\frac{1 - 
\theta_m^2}{2r_m}, \theta_m\right) = \left(\alpha', \theta' + \frac{2p}{\eta_m} 
\right)$, which indicates that the beams are focused on multiple angles in the 
distance ring $\alpha'$ in the near-field, as shown in Fig. \ref{img:rainbow} 
(a). 
Moreover, we can set $\theta'$ as an actual value and  $\alpha'$ as an abnormal 
value, then the beams are focused on multiple distances at the same angle as 
shown in Fig. \ref{img:rainbow} (b), 
satisfying $\left(\frac{1 - 
	\theta_m^2}{2r_m}, \theta_m\right) = \left(\alpha' + \frac{2q}{d\eta_m}, 
	\theta'\right)$.
As a result, by carefully designing the parameters $\alpha'$ and 
$\theta'$, the beams across the entire band could occupy multiple angles and 
distances simultaneously, and this property is more flexible than that in the 
far-field scenario.

{\color{black}
In conclusion, the mechanism of near-field controllable beam split indicates 
that a TD beamformer can 
not only mitigate the near-field beam split effect, but also control its 
degree, i.e., one can flexibly adjust the covered angular range and distance 
range of the beams generated by TD beamformer over the entire 
bandwidth. 
A simple analogy of this effect is the dispersion 
of white light caused by a prism. 
Since a prism has different refractive indices for the wideband white light,
the different frequency components of the pure light will disperse and 
eventually produce the rainbow. 
In our discussion, the function of time-delay circuits is similar to that of 
the prism. Therefore, the near-field controllable beam split effect is also 
termed as ``{near-field rainbow}" in this paper.
}
  
By elaborately designing the pattern of the near-field rainbow as shown in Fig. 
\ref{img:rainbow}, efficient near-field beam management can be 
realized, which will be discussed in the next section. 

%\text{}

	\section{Proposed Beam Training Scheme} \label{sec:4}
	 In this section, we first introduce the concept of near-field beam 
	 training and indicate the overhead for exhaustive near-field beam training 
	 is much larger than the far-field beam training, which is unacceptable in 
	 XL-MIMO systems. 
	To solve this problem, we propose a wideband beam training algorithm by 
	taking advantage of the near-field rainbow.
	\subsection{Exhaustive Near-Field Beam Training}\label{sec:4-1}
%	To realize efficient near-field beamforming, it is essential to obtain the 
%accurate channels. 
	Beam training is a well-adopted scheme to obtain CSI for current 5G systems
	\cite{BT_Wu2020}. 
	The concept of near-field beam training can be derived from the classical 
	far-field beam training.
	For the classical far-field beam training, the spatial angle information 
	$\theta_0$ is desired. The optimal beamsteering vector for a user is 
	selected from a predefined far-field beam codebook through a training 
	procedure between the BS and the user. Generally, each codeword in the 
	far-field codebook determines a unique spatial angle, where the distance is 
	assumed to be very large so that the near-field property is ignored. 
	Therefore, the entire far-field codebook occupies all of the potential 
	angles in the far-field.
	
	Similarly, the near-field beam training desires to obtain the location 
	information $(r_0, \theta_0)$ of the dominant path between the BS and the 
	user. The optimal beamfocusing vector is selected from a predefined 
	near-field beam codebook through a training procedure between the BS and 
	the user. Each near-field codeword determines a unique location, and the 
	entire near-field codebook occupies all of the desired angles and 
	distances. 	
	
	To be more specific, we introduce the procedure of exhaustive near-field 
	beam training realized by a TD beamformer. 
	Notice that the procedure below is also valid for the PS beamformer when 
	the 
	bandwidth is not very large. 
	We define $[\theta_{\min}, \theta_{\max}]$ as the potential range of 
	spatial angle, satisfying $-1 \le \theta_{\min} \le \theta_{\max} \le 1$. 
	Moreover, we assume the minimum distance between the BS and the user is 
	$\rho_{\min}$, so that the potential range of distance is $r \in 
	[\rho_{\min}, +\infty]$ or $\alpha = \frac{1 - \theta^2}{2 r} \in [0, 
	\alpha_{\max}]$ with $\alpha_{\max} = \frac{1}{2\rho_{\min}}$. Then, 
	multiple angles from $\theta \in [\theta_{\min}, \theta_{\max}]$ and  
	distances from  $r \in [\rho_{\min}, +\infty]$ are sampled simultaneously 
	to construct the near-field codebook. As proved in \cite{NearCE_Cui2022}, 
	the sampled locations could satisfy
	\begin{align}
	\theta_u &= \theta_{\min} + \frac{u}{U}(\theta_{\max} - \theta_{\min}),\label{eq:theta}\\
	\alpha_s &= \frac{s}{S}\alpha_{\max},   \label{eq:alpha}
	\end{align}
	where $u = 0,1,\cdots, U-1$ and $s = 0,1, \cdots, S-1$. $U$ denotes the 
	number of sampled angles and $S$ denotes the number of sampled distance 
	rings.
	Notice that when $S = 1$, we have $\alpha_s \equiv 0$, and then the 
	near-field codebook is naturally decayed to the far-field codebook.
	The exhaustive near-field beam training scheme  searches the entire 
	codebook 
	$\{\theta_u\}$ and $\{\alpha_s\}$ to obtain the optimal beamfocusing 
	vector. Apparently, the overhead for exhaustive near-field beam 
	training, i.e., the number of time slots used for beam training, is $T_1 = 
	US$. 
	
	In the $t$-th time slot, where $t = sU + u$ with $t = 0,1,\cdots, T_1-1$, 
	$u = 0,1,\cdots, U-1$, and $s = 0,1, \cdots, S-1$, we set the parameters of 
	the TD beamformer as $\theta_t' = \theta_u = \theta_{\min} + 
	\frac{u}{U}(\theta_{\max} - \theta_{\min})$ and $\alpha_t' = \alpha_s = 
	\frac{s}{S}\alpha_{\max}$, and then generate the beamfocusing vector 
	$\mb{w}_m(\theta_t', \alpha_t')$ according to (\ref{eq:TTD}). 
	Since $\theta_t'$ and $\alpha_t'$ are actual values, the BS is able to 
	transmit the pilot sequence to the user by the beam focused on the 
	location $(r_t', \theta_t') = \left( \frac{1 - \theta_u^2}{2\alpha_s}, 
	\theta_u\right)$. So the received signal $y_{m,t}$  in the $t$-th time slot 
	at $f_m$ is 
	\begin{align}\label{eq:y1}
		y_{m,t} = \sqrt{P_t}\mb{h}_m^T \mb{w}_m(\theta_t', \alpha_t') x_m + 
		n_m, 
	\end{align}
	where $n_m \sim \mathcal{CN}(0, \sigma^2)$ denotes the Gaussian noise, 
	$P_t$ 
	denotes the transmit power, and
	$x_m$ denotes the transmit pilot satisfying 
	$\|x_m\|^2 = 1$.
	After $T_1$ time slots, the estimated physical location $(\hat{r}, 
	\hat{\theta}) = (\frac{1 - \hat{\theta}^2}{2\hat\alpha}, \hat{\theta})$ 
	corresponding to the largest user received power can 
	be selected  from the $T_1$ measured locations, where,
	\begin{align}
	( \hat\theta, \hat{\alpha}) = \arg\max_{\theta_t', \alpha_t'}\sum_{m = 1}^{M} \|y_{m,t}\|^2.
	\end{align}
	Finally, a near-field beam $\mb{w}_m(\hat\theta, \hat{\alpha})$ aligned 
	with the location 
	$(\hat{r},\hat{\theta})$ can be generated to serve the user, and a  
	near-optimal beamfocusing gain can be achieved.
	
	Nevertheless, since the exhaustive near-field beam training scheme has to 
	search the entire near-field codebook exhaustively, and the scale of the 
	near-field codebook is much larger than that of the far-field, i.e. $US \gg
	U$,
	the training overhead is unacceptable in practice.
	Therefore, a near-field beam training scheme with low pilot overhead 
	 is essential for XL-MIMO systems.
		
%	Nevertheless, since it searches the training codebook exhaustively, the training overhead $T_1 = DS$ is so large that unaffordable in practice. 
%	For instance, if the BS antenna number is $N = 256$ and the carrier is $f_c = 100$ GHz, then $D$ is usually set as $D = N = 256$ while $S$ is generally set as 6 \cite{NearCE_Cui2022}. The training overhead is thus $T_1 = 1536$, which is much larger than the antenna number $N$, and thus is unacceptable in practice. Consequently, the solution to realize low-overhead near-field beam training is essential for XL-MIMO systems.
	
	\subsection{Proposed Near-Field Beam Training Scheme} \label{sec:4-2}
	The main reason for the high training overhead of exhaustive near-field 
	beam training is that only one physical location can be measured in each 
	time slot. By contrast, as we discussed in Section \ref{sec:3-2}, a 
	TD beamformer is able to generate multiple beams focusing on multiple 
	locations by 
	only 
	one RF chain. 
 	Taking advantage of the near-field rainbow, multiple physical locations can 
 	be measured simultaneously in each time slot. 
 	Inspired by this observation, we propose a near-field rainbow-based beam 
 	training method to 
 		significantly reduce the training overhead.
 		
 	Generally, since the spatial resolution of an antenna array on the angle is 
 	much higher than that on the distance, the number of sampled angles $U$ is 
 	usually much larger than the number of sampled distances $S$ 
 	\cite{NearCE_Cui2022}. 
 	Thus, the training overhead is mainly determined by the search of angle 
 	$\theta_0$. 
 	Therefore, the near-field rainbow on the angle dimension, as shown in Fig. 
 	\ref{img:rainbow} (a), is utilized to avoid the exhaustive search of 
 	angle, i.e., the angle-related parameters $\theta'$ are set as abnormal 
 	values, while the distance-related parameters $\alpha'$ 
 	can be set as actual values. 
 	In other words, the proposed scheme searches 
 	the optimal angle in a frequency division manner, and searches the optimal 
 	distance ring in a time division manner.  The specific procedure  of 
 	the proposed beam training scheme  is illustrated in \textbf{Algorithm 1}. 
% 			\begin{figure}
% 		\centering
% 		%		\vspace*{-0.5em}
% 		\includegraphics[width=3.5in]{figure/rainbow.pdf}
% 		\caption{  The schematic diagram for the NF-CBS-based beam training.
% 		} 
% 		\label{img:rainbow}
% 		\vspace*{-1em}
% 	\end{figure}

\begin{algorithm}[htb]
	\color{black}
	\caption{$\!\!$:  Proposed near-field rainbow based beam 
	training 
	scheme}
	\label{alg1}
	\begin{algorithmic}[1]
		\REQUIRE ~~\\
		Potential angular range $[\theta_{\min}, \theta_{\max}]$, focused angle 
		$\theta_c$ at the central frequency, potential distance range $[0, 
		\alpha_{\max}]$, the number of sampled distances $S$, central frequency 
		$f_c$, bandwidth $B$.
		\ENSURE ~~\\
		Estimated physical location $(\hat{r}, \hat{\theta})$\\
		\STATE
		$f_L = f_c - \frac{B}{2}$, $f_H = f_c + \frac{B}{2}$ \\
		\STATE
		Calculate the abnormal delay parameter: \\
		$\theta' = \theta_c 
		-2\left\lceil \max\left\{ f_L(\theta_{\max} - 
		\theta_c), f_H(\theta_c - \theta_{\min})\right\}/B\right\rceil$\\		
		\FOR {$t \in \{0,1,\cdots,{S-1}\}$}
		\STATE 
		Calculate the normal delay parameter: $\alpha_t' = 
		\frac{t}{S}\alpha_{\max}$\\
		\STATE
		Obtain the beamfocusing vector: $[\mb{w}_m(\theta', \alpha'_t)]_n  
		=\frac{1}{\sqrt{N_t}}  e^{-jk_m 
		(nd\theta' - 
			n^2 
			d^2 \alpha'_t )}$
		\\
		\STATE
		Received signal: $y_{m,t} = \sqrt{P_t}\mb{h}_m^T \mb{w}_m(\theta', 
		\alpha_t') x_m 
		+ n_m$\\
		\ENDFOR
		\STATE 
		$( \hat{m}, \hat{t}) = \arg\max_{m, t} 
		\|f_my_{m,t}\|^2,$\\
		\STATE
		$\hat{\theta} = \theta' + (\theta_c - \theta')f_c/f_{\hat{m}}$
		\STATE
		$\hat{\alpha} = \alpha'_{\hat{t}} $	
		\RETURN $(\hat{r}, \hat{\theta}) = \left(\frac{1 - 
		\hat{\theta}^2}{2\hat{\alpha}}, \hat{\theta}\right)$.
	\end{algorithmic}
\end{algorithm}

 	Firstly, in steps 1-2, we desire to design the abnormal parameter $\theta' 
 	\notin [-1, 1]$, so that the frequency-dependent angles $\theta_m = \theta' 
 	+ (2pf_c)/f_m$ across the entire band is able to cover the entire potential 
 	angle range $[\theta_{\min}, \theta_{\max}]$, as shown in Fig. 
 	\ref{img:rainbow} (a).
 	To realize this target, we first assume that, at the central frequency 
 	$f_c$, the 
 	beam is aligned with the angle $\theta_c = \theta' + 2p$, where $\theta_c$ 
 	is a predefined spatial angle satisfying $-1 \le \theta_{\min} < \theta_c < 
 	\theta_{\max} \le 1$. Then we have \begin{align}\label{eq:c1}
 	\theta' = \theta_c - 2p, \quad \:p \in \mathbb{Z}.
 	\end{align}
 	Without loss of generality, we assume the delay parameter $\theta' < -1$, 
 	then $p = 
 	\frac{\theta_c - \theta'}{2} > \frac{-1 - \theta'}{2} > 0$ is a positive 
 	integer. 
 	In this case, the frequency-dependent angle 
 	$\theta_m = \theta' 
 	+ (2pf_c)/f_m$ is monotonically decreasing with respect to the frequency
 	 $f_m$.
 	
 	Moreover, since the available bandwidth is $B$, the lowest frequency and 
 	the highest frequency are $f_L = f_c - \frac{B}{2}$ and $f_H = f_c + 
 	\frac{B}{2}$, respectively. 
 	Therefore, the minimum and maximum values of $\theta_m$ are $\theta' + 
 	\frac{2pf_c}{f_H}$ and $\theta' + \frac{2pf_c}{f_L}$, respectively.
 	To cover the entire potential angles $[\theta_{\min}, \theta_{\max}]$, it should satisfy 
 	\begin{align}
 	\theta_{\min} &\ge \theta' + (2pf_c)/f_H, \label{eq:c2}\\
 	\theta_{\max} &\le \theta' + (2pf_c)/f_L. \label{eq:c3}
 	\end{align}
 	By solving (\ref{eq:c1}), (\ref{eq:c2}), and (\ref{eq:c3}), one of the 
 	solutions to $\theta'$ is  
 	\begin{align}\label{eq:xi}
 	\theta' =  \theta_c -2\left\lceil \max\left\{ f_L(\theta_{\max} - \theta_c), f_H(\theta_c - \theta_{\min})\right\}/B\right\rceil,
 	\end{align}
 	where $\lceil x \rceil$ denotes the nearest integer greater than or equal 
 	to $x$, and $p$ is given by $p = \frac{\theta_c -\theta'}{2} = \left\lceil 
 	\frac{1}{B}\max\left\{ f_L(\theta_{\max} - \theta_c), f_H(\theta_c - 
 	\theta_{\min})\right\}\right\rceil$.
 	
 	Then, in step 4, different distance rings are searched in different time 
 	slots  	
 	by adjusting the normal delay 
 	parameter $\alpha'_t$.
 	Similar to the exhaustive near-field beam training scheme, the potential 
 	range of distance ring is $\alpha \in [0, \alpha_{\max}]$, and the number 
 	of distance rings to be measured is $S$.
 	Therefore, in the $t$-th time slot, $\alpha_t'$ is set as 
 	$\frac{t}{S}\alpha_{\max}$ with $t = 0,1,\cdots, S - 1$ to 
 	 	measure the beamfocusing gains on the distance ring $\frac{1 - 
 	 	\theta^2}{2r} = 
 	 	\alpha_t'$. 
 	 	
 	 {\color{black}	
 	As a result, with the parameters $\alpha_t'$ and $\theta'$, as shown in 
 	Fig. \ref{img:rainbow} (a), the beams $\mb{w}_m(\theta', \alpha_t')$ 
 	generated 
 	in step 5 by the TD beamformer is able to occupy the entire angular range 
 	in the 
 	distance ring $\alpha_t'$.
 	Utilizing these beams, in step 6, the received signal $y_{m,t}$  in the 
 	$t$-th time slot at frequency $f_m$ is denoted as
 	\begin{align}\label{eq:y2}
 	y_{m,t} & = \sqrt{P_t}\mb{h}_m^T \mb{w}_m(\theta', \alpha_t') x_m + n_m 
 	\notag \\
 	& = \sqrt{P_t}\sqrt{N_t}  \beta_m  \mb{a}_m^T(\theta_0, 
 	r_0)\mb{w}_m(\theta', \alpha_t')x_m  + n_m \notag \\
 	& = \sqrt{P_tN_t}  \beta_m  g_{m,t}x_m  + n_m,
 	\end{align}
 	where $g_{m,t} = \mb{a}_m^T(\theta_0, 
 	r_0)\mb{w}_m(\theta', \alpha_t')$.
 	The optimal beamfocusing vector is corresponding to the maximum 
 	beamfocusing gain $\|g_{m,t}\|^2$. Since $\|x_m\|^2 = 1$ and $n_m \sim 
 	\mathcal{CN}(0, \sigma^2)$, the maximum likelihood estimation of $g_{m,t}$ 
 	is 
 	\begin{align}\label{eq:g}
 	\hat{g}_{m,t} = \frac{y_{m,t}}{\sqrt{P_tN_t}  \beta_m x_m}.
 	\end{align}
 	Thus, the beamfocusing gain $\|\hat{g}_{m,t}\|^2$ can be estimated as 
 	$\|\hat{g}_{m,t}\|^2 = \frac{1}{P_tN_t }\frac{\|y_{m,t}\|^2}{ 
 	\|\beta_m\|^2}$. Based on the definition of $\|\beta_m\|^2$ in Section 
 	\ref{sec:2}, we have 
 	$\|\hat{g}_{m,t}\|^2 = C \|f_my_{m,t}\|^2$, where $C$ is a constant that is 
 	not relevant to $f_m$. 
 	Notice that the proposed beam training scheme does not require the specific 
 	value of ${\|\beta_{m}\|^2}$. As long as the relationship between  
 	$\beta_{m}$ and $f_m$ is available, whether quadratic function, 
 	exponential function, and so on, the proposed scheme is always valid.
 }
 	
 	Besides, in step 8, based on the formulation of beamfocusing gain 
 	$\|\hat{g}_{m,t}\|^2 = C \|f_my_{m,t}\|^2$, the label of the optimal beam 
 	$\hat{t}$ and $\hat{m}$ is acquired by maximizing $\|\hat{g}_{m,t}\|^2$ 
 	over the total $S$ measured distance rings and $M$ subcarriers, where  
 	\begin{align}
 	( \hat{m}, \hat{t}) = \arg\max_{m, t} \|f_my_{m,t}\|^2.
 	\end{align}

Eventually, in steps 9-10, based on the mechanism of the near-field  rainbow 
(\ref{eq:theta2}), 
the estimated spatial location is 
 	\begin{align}
 	%	\hat\theta &= \theta' +  \frac{2p}{\eta_{\hat{m}}} = \xi +  \frac{\theta_c - \xi}{\eta_{\hat{m}}}. \\
 	\hat\theta &= \theta' + 2p/\eta_{\hat{m}} = \theta' +  {(\theta_c - \theta')f_c}/{f_{\hat{m}}}, \\
 	\hat\alpha &= \alpha_{\hat t}'.
 	\end{align}
 	After that, the beam training is completed, and 
 	the BS can generate the beamfocusing vector 
 	$\mb{w}_m(\hat\theta, \hat\alpha)$ to serve the user for data transmission.

	The advantage of the proposed near-field rainbow based beam training method 
	is that the 
	optimal angle is searched out in a frequency division manner. Therefore, 
	the training overhead is only determined by the overhead for searching the 
	optimal distance ring, which is significantly reduced.
	In the next sub-section, we quantitatively analyze the training overheads of the two near-field beam training schemes above.
	
	\subsection{Comparison on the Beam Training Overhead}
	Beam training overhead refers to the number of time slots used for beam 
	training. It is obvious that, the training overhead of the exhaustive 
	near-field beam training scheme is $T_1 = US$, while the training overhead 
	of the proposed near-field rainbow based scheme is $T_2 = S$.  
	As discussed in \cite{NearCE_Cui2022}, the number of distance rings is 
	generally much less than the number of angles.
	For instance, if the BS antenna number is $N = 256$, the carrier is $f_c = 60$ GHz, and the minimum distance is $\rho_{\min} = 2$ m, then $U$ is usually set as $U = N = 256$ while $S$ is generally set as 10 \cite{NearCE_Cui2022}.
	In this case, the training overhead $T_2 = S = 10$ is much less than $T_1 = 
	US = 2560$. Therefore, benefiting from the near-field rainbow, the 
	proposed scheme 
	is able to significantly reduce the pilot overhead for near-field beam 
	training, which will be further verified by simulation results in Section V.

	%	Let the parameters of $\theta'$ and $\alpha'$ traverse all of .

	\section{Simulation Results} \label{sec:5}
In this section, simulations are provided to verify the effect of
the near-field rainbow and demonstrate the performance of the proposed 
near-field 
rainbow based beam training 
scheme. We consider a wideband XL-MIMO system, where the BS equips an $N_t = 
256$-element ULA with a TD beamformer. The carrier frequency is $f_c = 60$ GHz, 
the 
number of sub-carriers is $M = 2048$, and the bandwidth is $B = 3$ GHz. 
%For the high frequency channel, considering that the signals are quasi-optimal 
%and the 
%channel is LoS-dominant, we set $L = 1$. The path gain is considered to be 
%frequency-dependent with the model in (\ref{eq:fspl2}). We assume the path 
%gain 
%at the center frequency $\left|\beta^{(0)}_c\right|^2$ as the reference path
%gain with $\left|\beta^{(0)}_c\right|^2 = 1$ without loss of generality.

\subsection{The Demonstration of Near-Field Controllable Beam Split}
	\begin{figure}
		\centering
		%		\vspace*{-0.5em}
	\subfigure[]
{\includegraphics[width=3.2in]{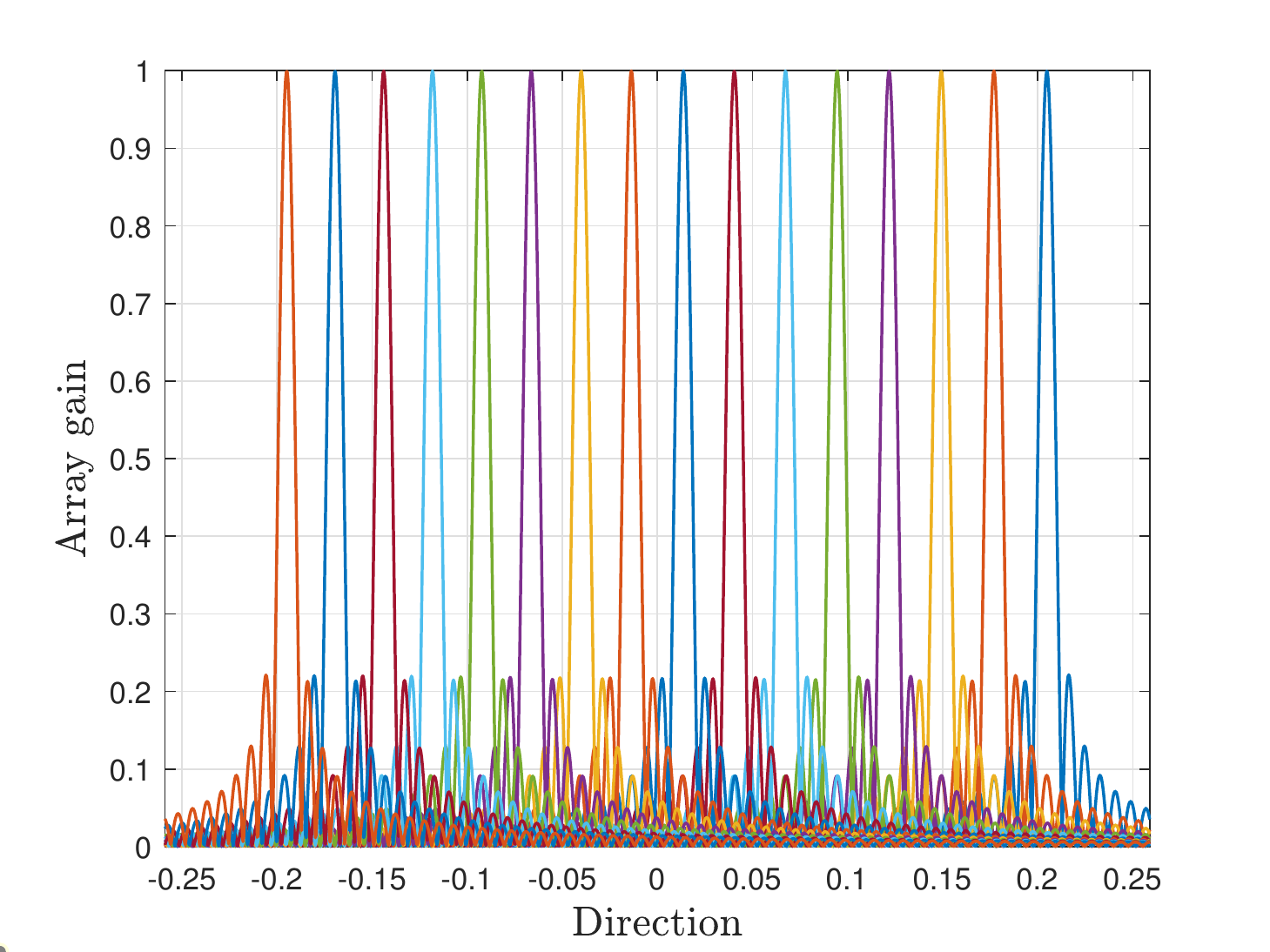}}
\subfigure[]
{\includegraphics[width=3.2in]{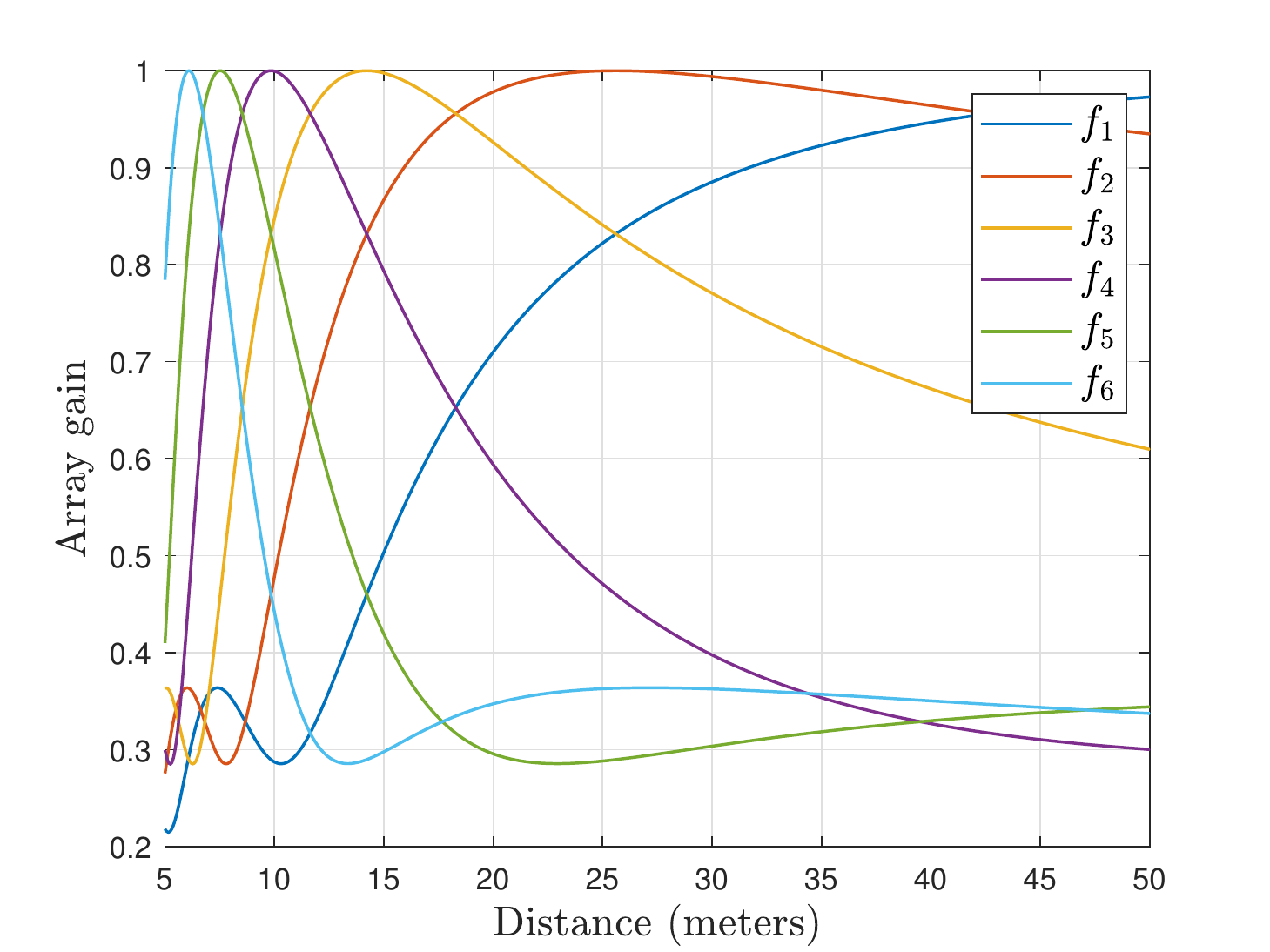}} 
		\caption{  Near-field rainbow achieved by a TD beamformer. In (a), we 
		set 
		$\theta' = -6 \notin[-1, 1]$ and $\alpha' = \frac{1}{20} \text{m}^{-1} 
		> 0$. The array gains at multiple frequencies w.r.t the angle on 
		distance ring $\alpha'$ are evaluated. In (b), we set $\theta' = 
		\frac{\pi}{8} \in[-1, 1]$ and $\alpha' = -\frac{2}{d} < 0$. The array 
		gains at multiple frequencies w.r.t the distance on angle $\theta'$ are 
		evaluated. 
		} 
		\label{sim:rainbow}
		\vspace*{-1em}
	\end{figure}
Firstly, in Fig. \ref{sim:rainbow}, we verify that time-delay circuits are able 
to produce near-field controllable beam split.
Specifically, in Fig. \ref{sim:rainbow} (a), the near-field rainbow on the 
dimension of 
angle is evaluated. We set the delay parameter $\theta'$ as $\theta' = -6 
\notin [-1, 1]$ 
and the parameter $\alpha'$ as $\alpha' = \frac{1}{2r'}$ with $r' = 10$ m. 
On the distance ring $\alpha' = \frac{1 - \theta^2}{2r}$, the array gains at 
different frequencies with respect to (w.r.t) the angle are shown in Fig. 
\ref{sim:rainbow} (a).
For clarity, only a few frequencies are plotted. The beams over different 
frequencies are focused on multiple angles in the distance ring $\alpha' = 
\frac{1 - \theta^2}{2r}$, and cover a given angular range $[-0.2, 0.2]$. 
Therefore, the near-field rainbow on the angle dimension is verified.

Then, in Fig. \ref{sim:rainbow} (b), the near-field rainbow on the dimension of 
distance is 
evaluated. We set the angle $\theta'$ as $\theta' = \sin\frac{\pi}{8} \in [-1, 
1]$ and the distance ring $\alpha'$ as $\alpha' = -\frac{2}{d} < 0$. In the 
physical angle $\theta'$, the array gains at different frequencies w.s.t the 
distance are shown in Fig. \ref{sim:rainbow}(b).
For clarity, only a few frequencies are plotted. The beams over different 
frequencies are focused on multiple distances in the angle $\theta'$, which 
covers the entire distance range.
Therefore, the near-field rainbow on the distance dimension is also achievable.

\subsection{Beam Training Performance}

In this subsection, the performance of the proposed near-field rainbow based 
beam training scheme is evaluated. The potential spatial angle range of the 
user is set as $[\theta_{\min}, 
\theta_{\max}] = [-\sin\frac{\pi}{3}, \sin\frac{\pi}{3}]$, while the 
potential distance of the user is set as $[\rho_{\min}, +\infty] = [3 \: 
\text{m}, 
+\infty]$ and thus the range of $\alpha$ is $\alpha \in [0, \frac{1}{6} 
\text{m}^{-1}]$. The spatial angle at the center frequency is fixed to 
$\theta_c 
= 0$. Then, for the proposed near-field rainbow based scheme, according to 
(\ref{eq:xi}), the delay parameter
$\theta'$ is acquired as $\theta' = \xi = -36$ with $p = 18$. 
Moreover, we set the number of 
distance rings to be searched as $S = 10$. For the exhaustive near-field beam 
training, the number of angles to be searched is $U = N = 256$. Finally, 
we use the average rate performance to quantify the beam training performance, 
which is mathematically defined as {\color{black}
\begin{align}
R &= \frac{1}{M}\sum_{m=1}^{M}\log_2\left(1 + 
\frac{P_t}{\sigma^2}\|\mb{h}_m^T \mb{w}_m\|^2\right) = 
\frac{1}{M}\sum_{m=1}^{M}\log_2\left(1 + 
\frac{P_tN_t\beta_m^2}{\sigma^2}\|\mb{a}_m^T(\theta_0, r_0) \mb{w}_m\|^2\right),
\end{align}
where $\mb{w}_m$ denotes the beamfocusing vector for data transmission searched 
by beam training and SNR $=\frac{P_tN_t\beta_m^2}{\sigma^2}$ denotes the 
signal-to-power ratio. }The 
compared benchmarks are shown below.
\begin{itemize}
	\item Perfect CSI: The perfect channel $\mb{h}_m$ is available at the BS, 
	which is served as the benchmark for the upper bound of average rate 
	performance.
	\item Far-field rainbow based beam training: The classical wideband 
	far-field beam training scheme achieved by TD beamformer in 
	\cite{TTDBT_Bol2021}. This method can be regarded as the far-field 
	rainbow-based beam training scheme, where only the optimal angle is 
	searched in a frequency-division manner and the distance ring $\alpha$ is 
	assumed to be zero.  
	{\color{black}\item Near-field hierarchical beam training: This method is 
	the 
	hierarchical beam scheme training designed for reconfigurable intelligence 
	surface (RIS) aided near-field
	communications \cite{NFBT_Hong2021}. We transform this method to XL-MIMO 
	scenarios for comparison.
	\item Far-field hierarchical beam training: This method is the classical 
	hierarchical beam 
	training designed for the far-field scenarios \cite{BT_Noh2017}.}
	\item Exhaustive search: This method is the exhaustive near-field beam 
	training scheme provided in Section \ref{sec:4-1}. In the training 
	procedure, we first fix the distance ring $\alpha_s$ and exhaustively 
	search all of the angles 
	$\theta_u$. Then, we change the distance ring $\alpha_s$ and repeat the 
	above steps until all of the distance rings and angles are measured.  
\end{itemize}

\begin{figure}
	\centering
	\subfigure[]
	{\includegraphics[width=3.2in]{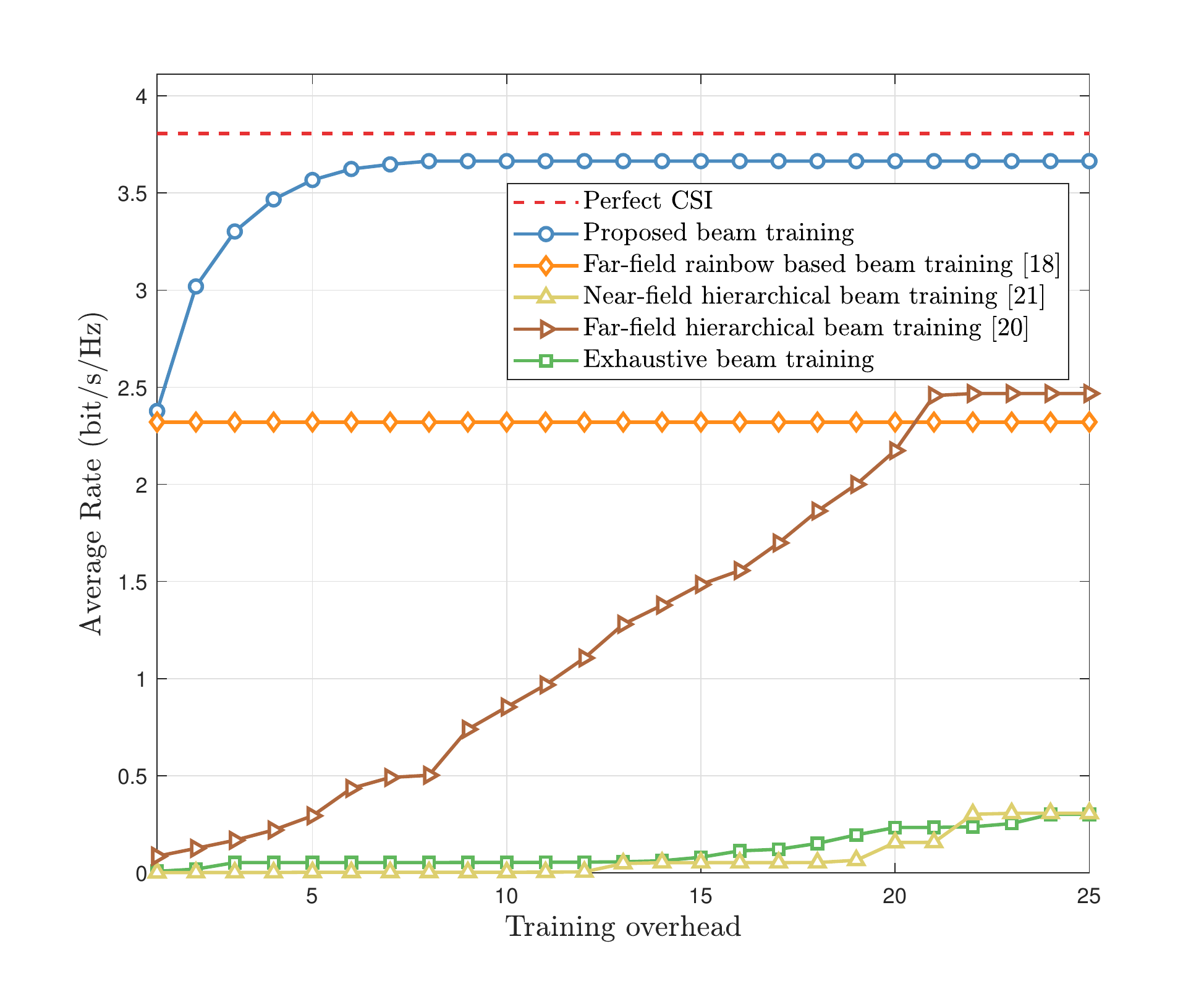}}
	\subfigure[]
	{\includegraphics[width=3.2in]{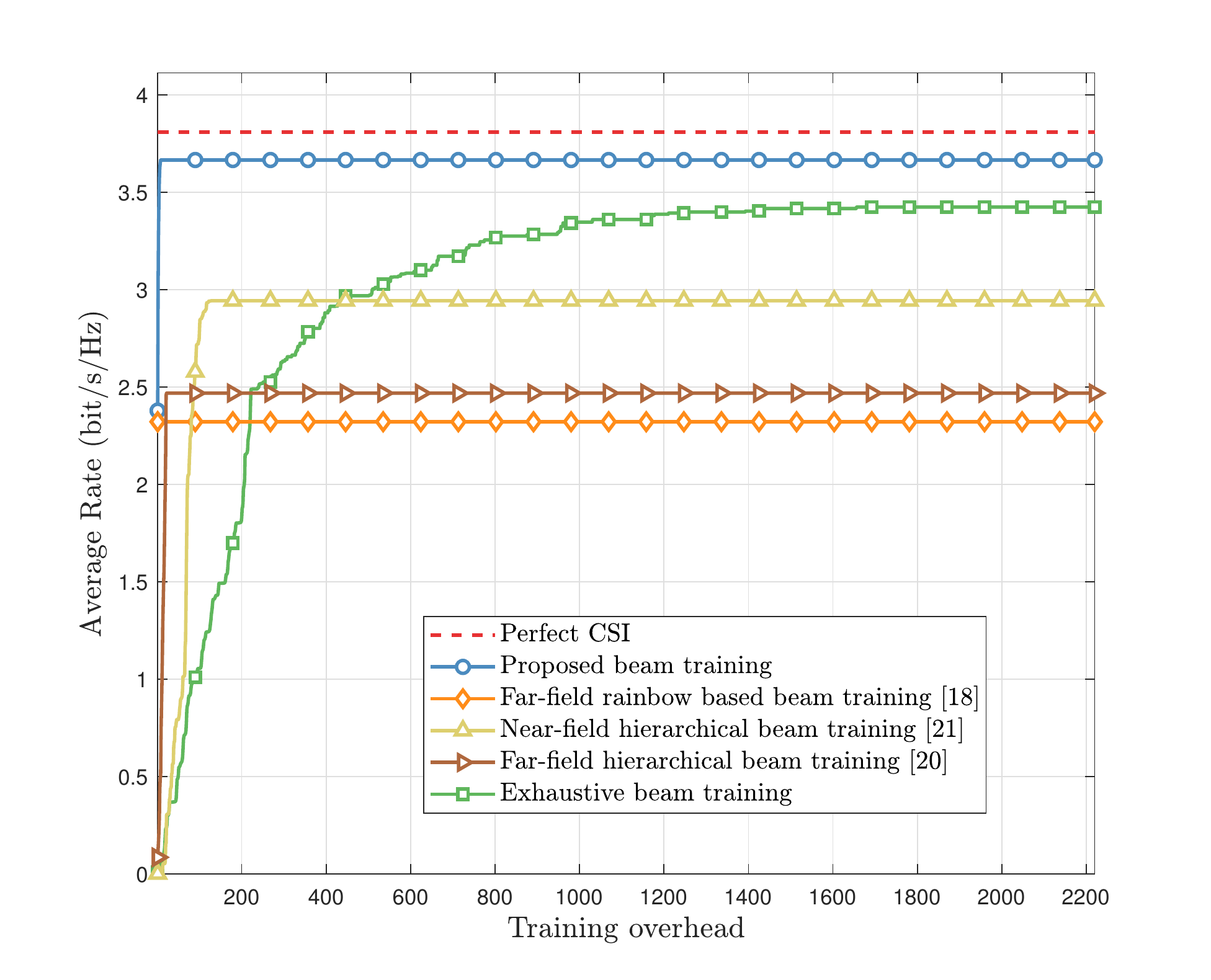}} 
	\\ %换行
	\centering
	\caption{ \color{black} Average rate  performance vs. training overhead.
	} %图片标题
	\label{sim:Rate_overhead_5}
\end{figure}
{\color{black}
Fig. \ref{sim:Rate_overhead_5} illustrates the average rate performance against 
the training overhead. The training overhead is increasing from 0 to $US = 
2560$. The SNR is 
set as 10 dB. 
$10^3$ realizations of the user location are generated for Monte Carlo 
simulations, where  $\theta_0 \sim 
\mathcal{U}(-\sin\frac{\pi}{3}, \sin\frac{\pi}{3})$ and $r_0 \sim 
\mathcal{U}(3\:\text{meters}, 30\:\text{meters})$. In the $t$-th time slot, we 
utilize the optimal beamforming vector searched during the time slots $1 \sim 
t$ to serve the user. Specifically, the training overhead of the two far-field 
schemes are both very low, e.g., 22 overhead for the hierarchical far-field 
scheme and only one overhead for the far-field rainbow based scheme. 
However, since the far-field schemes only 
consider the angle information while ignoring the distance information, their 
average rate performance is not satisfactory. Besides, to achieve a 
satisfactory average rate, the training overhead of the near-field hierarchical 
scheme and the exhaustive scheme will be very high, e.g., 200 overhead for the 
near-field hierarchical scheme and more than 1000 for the exhaustive scheme. By 
contrast, the proposed near-field rainbow based beam training scheme enjoys a 
much higher average rate performance with very low overhead. This is thanks 
to two factors: 1) both the angle and distance information are considered, 2) 
the near-field rainbow effect is exploited to avoid the exhaustive search for 
the angle information. Actually, the proposed scheme has already achieved 96\% 
of the average rate benchmark with only 8 training overhead. 
}

\begin{figure}
	\centering
	%		\vspace*{-0.5em}
	\includegraphics[width=3.5in]{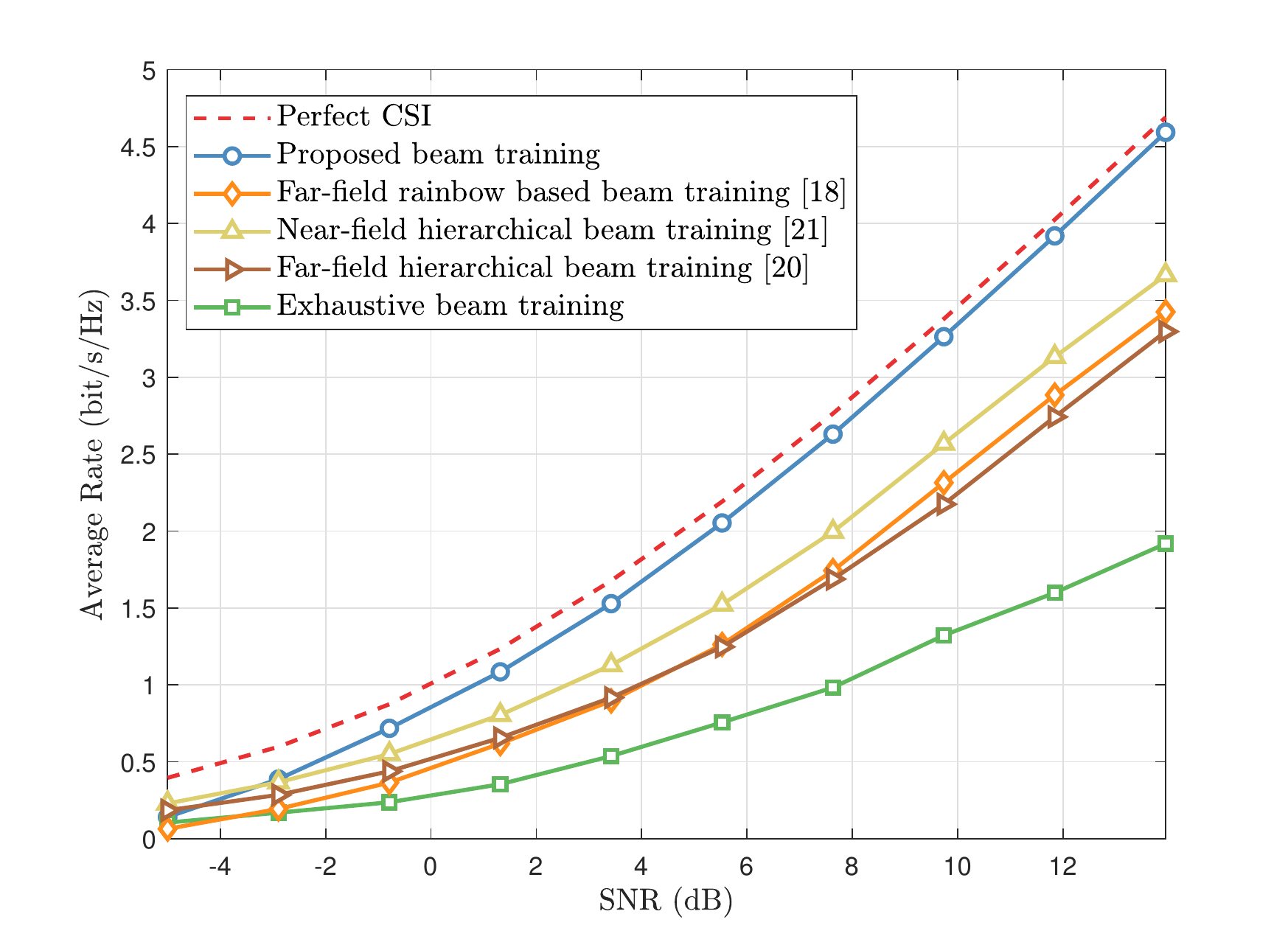}
	\caption{ \color{black} Average rate performance vs. SNR.
	} 
	\label{sim:rate_snr}
	\vspace*{-1em}
\end{figure}

\begin{figure}
	\centering
	%		\vspace*{-0.5em}
	\includegraphics[width=3.5in]{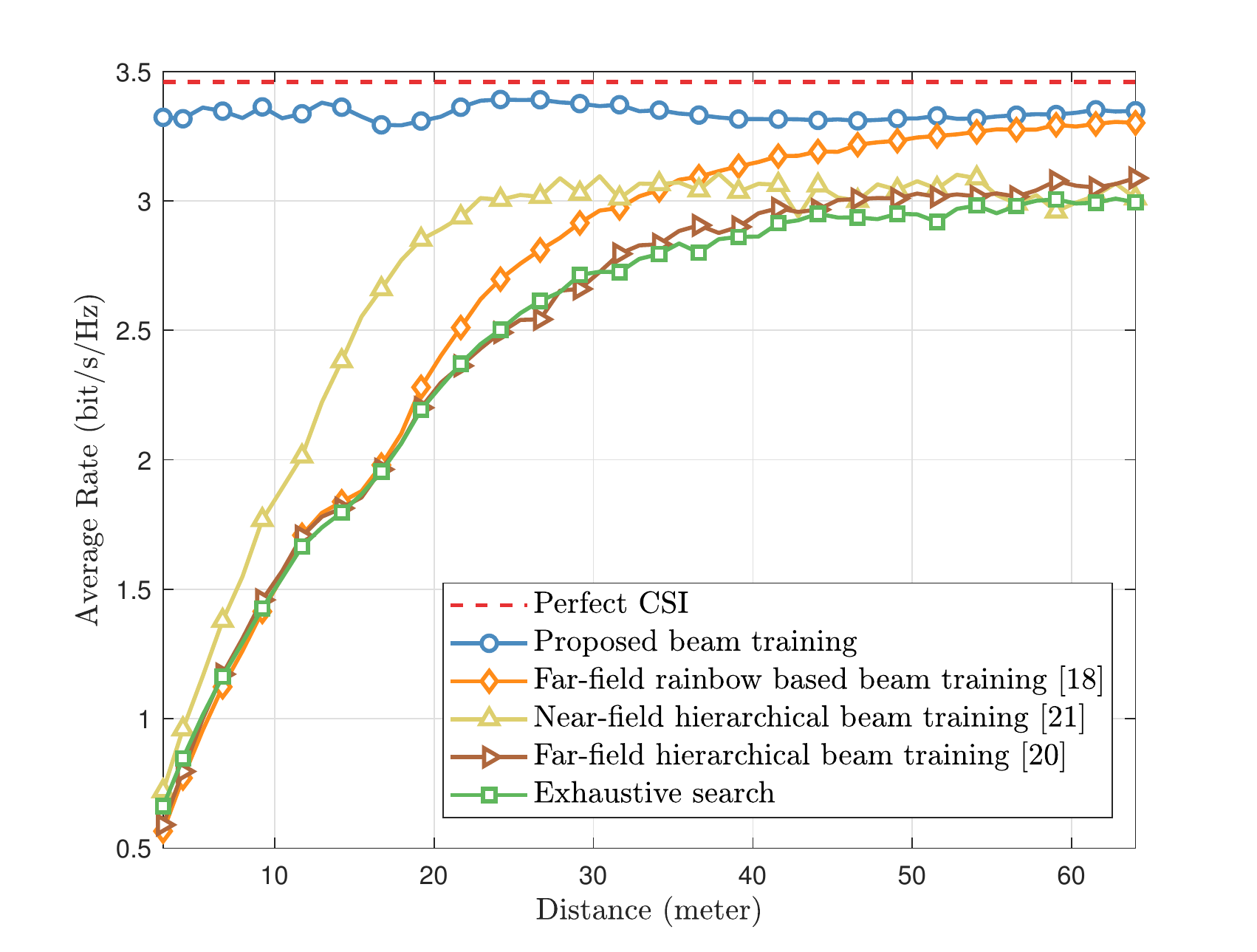}
	\caption{\color{black}  Average rate performance vs. distance.
	} 
	\label{sim:rate_distance_pi6}
	\vspace*{-1em}
\end{figure}

Fig. \ref{sim:rate_snr} shows the impact of SNR on different beam training 
schemes. Here, the SNR is increasing from -5 dB to 15 dB.
The maximum training overhead $T_{\max}$ for all considered methods is 
set as $T_{\max} = 256$. Notice 
that in our simulations, only the overhead for exhaustive beam training is 
$T_{\max} = 256$, while the overhead for other schemes is set according to 
their overhead requirements as illustrated in Fig. \ref{sim:Rate_overhead_5}, 
which are always less than $T_{\max}$. 
 The other simulation settings are the same as those in Fig. 
 \ref{sim:Rate_overhead_5}. 
It is clear from Fig. \ref{sim:rate_snr} that, the proposed scheme outperforms 
all existing far-field and near-field beam training schemes, and is able to 
achieve the near-optimal achievable average rate performance compared with the 
benchmark.
In addition, we can observe that the exhaustive beam training scheme suffers 
from severe degradation of the average rate performance. This is mainly 
because the maximum 
training overhead (256) is much less than the required training overhead (more 
than 1000) for the exhaustive scheme to achieve satisfactory average rate 
performance.
%more than 95\% the optimal average rate with much low training overhead.

In Fig. \ref{sim:rate_distance_pi6}, we illustrate the average rate performance 
against the distance. 
Here, the distance $r_0$ between the user and BS is gradually increasing from 3 
meters to 60 meters. The parameters are set as follows: $\text{SNR} = 10$ dB, 
$T_{\max} = 256$, $\theta_0 \in \mathcal{U}(-\sin\frac{\pi}{6}, 
\sin\frac{\pi}{6})$. 
 The other simulation settings are the same as those in Fig. 
\ref{sim:rate_snr}. 
The impact of near-field propagation on the beam training performance is clear 
in Fig. \ref{sim:rate_distance_pi6}.
For the far-field schemes, with the decrease of distance, near-field 
propagation becomes dominant and thus the average rate of these far-field 
schemes rapidly deteriorates. As for the degradation of the exhaustive scheme, 
since the maximum training overhead is limited to 256, this scheme is hard to 
search the locations in the near-field. Moreover, although the near-field 
hierarchical beam training scheme is able to slightly alleviate the average 
rate loss, its performance is unacceptable when the distance is less than 20 
meters.
By contrast, since the proposed scheme is able to search the optimal angle and 
distance with very low pilot overhead by exploiting the near-field rainbow, its 
performance is robust to all considered distances, whether in the near-field or 
in the far-field.

\begin{figure}
	\centering
	%		\vspace*{-0.5em}
	\includegraphics[width=3.5in]{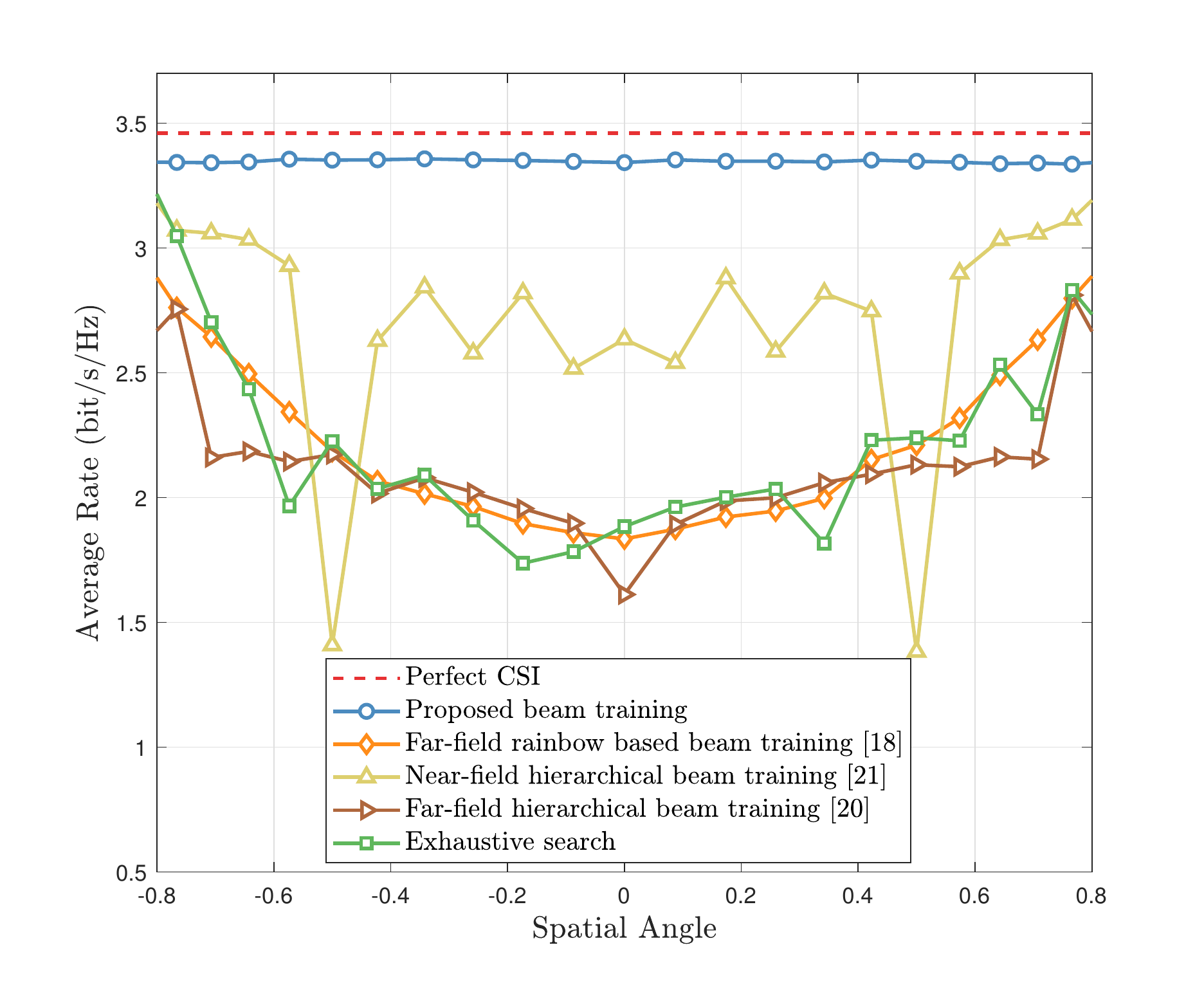}
	\caption{ \color{black} Average rate performance vs. angle.
	} 
	\label{sim:rate_angle_10m}
	\vspace*{-1em}
\end{figure}

{\color{black}
Fig. \ref{sim:rate_angle_10m} shows the average rate performance against the 
angle $\theta_0$. 
Here, the angle $\theta_0$ is gradually increasing from $-\sin\frac{\pi}{3}$ to 
$\sin\frac{\pi}{3}$ and the distance $r_0$ is randomly generated from $r_0 \sim 
\mathcal{U}(3\:\text{meters}, 30\:\text{meters})$. The other simulation 
settings are the same as those in Fig. \ref{sim:rate_distance_pi6}.
Notice that although the distance $r_0$ 
is fixed, the distance ring $\alpha_0 = \frac{1 - \theta_0^2}{2r_0}$ still 
varies with the angle $\theta_0$. 
For the far-field schemes, their performance is severely degraded when the 
angle $\theta_0$ is around zero. This is because the near-field property is 
more significant when the angle is around zero, which has been proved in 
\cite{NFBF_Cui2021}. 
Moreover, there exists severe fluctuation for the near-field hierarchical 
beam training scheme. 
This is because the near-field hierarchical scheme 
proposed in \cite{NFBT_Hong2021} creates the near-field codebook by uniformly  
sampling the codeword in the cartesian coordinates. It has been indicated in 
\cite{NearCE_Cui2022} that this kind of codebook cannot realize satisfactory 
beam training performance in the entire near-field environment.
By contrast, the proposed scheme achieves a near-optimal average rate for all 
considered angles.

}

	\section{Conclusions} \label{sec:6}
	This paper investigated the wideband near-field beam training for XL-MIMO 
	systems. The mechanism of the near-field rainbow is revealed, i.e., 
	time-delay 
	circuits can flexibly control the degree of the near-field beam split 
	effect. 
	Then, a near-field rainbow based beam training scheme was proposed to 
	search the optimal angle in a frequency-division manner, and the optimal 
	distance is 
	searched in a time-division manner.  
	The simulation results verified that: i) the beams generated by a 
	TD beamformer over the entire bandwidth can cover multiple angles and 
	distances,  ii) our near-field rainbow based scheme can achieve a 
	near-optimal average 
	rate with much-reduced training overhead, iii) the performance of our 
	scheme is robust to the distance and angle. 
	This paper unveiled that although the near-field beam split effect induces 
	a severe beamfocusing gain loss, it also provides a new possibility to 
	realize fast near-field CSI acquisition.
	In our future work, we will extend the mechanism of the near-field rainbow  
	to 
	other beamforming architectures, such as the delay-phase precoding 
	architecture 
	\cite{DPP_Tan2019}.

	\section*{Appendix A. The Periodicity of $G(x, y)$}
	$G(x, y)$ is a periodic function against the vector variable $(x, y)$ with period $(\frac{2\pi}{d}, \frac{2\pi}{d^2})$. Specifically, for any integers $p \in \mathbb{Z}$ and $q\in \mathbb{Z}$, we have
\begin{align}
G\left(x - \frac{2p\pi}{d}, y - \frac{2q\pi}{d^2}\right) &= 
\frac{1}{N_t} \left|\sum_{n = -N}^N  e^{ jn d (x - \frac{2p\pi}{d}) - 
jn^2 d^2 (y - \frac{2q\pi}{d^2})} \right| \notag \\
&=\frac{1}{N_t} \left|\sum_{n = -N}^N  e^{ jn dx  - jn^2 d^2 y }  e^{j2\pi n(qn 
- p)} \right|. 
\end{align}
Since $2\pi n(qn - p)$ is an integer multiple of $2\pi$, we have $e^{j2\pi n(qn 
	- p)} = 1$. Therefore, 
\begin{align}
G\left(x - \frac{2p\pi}{d}, y - \frac{2q\pi}{d^2}\right) 
=\frac{1}{N_t} \left|\sum_{n = -N}^N  e^{ jn dx  - jn^2 d^2 y } \right| = G(x, 
y).
\end{align}
As a result, the periodicity of function $G(x, y)$ is proved.
%
%\section*{Appendix B. Derivation of the solution to $\theta'$}
%From (\ref{eq:c1}) we have
%\begin{align} \label{eq:apb1}
%\theta' = \theta_c - 2p.
%\end{align}
%Substituting (\ref{eq:apb1}) into (\ref{eq:c2}) and (\ref{eq:c3}), we have 
%$\theta_{\min} \ge \theta_c - 2p + \frac{2pf_c}{f_H}$ and $\theta_{\max} \le \theta_c - 2p + \frac{2pf_c}{f_L}$. Then $p$ is given by 
%\begin{align} \label{eq:apb2}
%p \ge  \frac{1}{B}\max\left\{ f_L(\theta_{\max} - \theta_c), f_H(\theta_c - \theta_{\min})\right\}.
%\end{align}
%Since $p$ is a positive integer, one of the solutions to $p$ is $p = \left\lceil \frac{1}{B}\max\left\{ f_L(\theta_{\max} - \theta_c), f_H(\theta_c - \theta_{\min})\right\}\right\rceil$. Denote $\xi$ as the corresponding solution to $\theta'$, therefore, we have 
%\begin{align}
%\xi = \theta_c - 2p = \theta_c -2\left\lceil \frac{1}{B}\max\left\{ f_L(\theta_{\max} - \theta_c), f_H(\theta_c - \theta_{\min})\right\}\right\rceil.
%\end{align}
	\balance
	\bibliographystyle{IEEEtran}
	\bibliography{IEEEabrv,refs}
	% that's all folks
\end{document}